\documentclass[apj]{emulateapj}

\usepackage{natbib}
\usepackage{xspace}
\def\hi{\textsc{Hi}\xspace}

\newcommand{\hii}{{H{\scriptsize II}}\xspace}
\newcommand{\IRAS}{{\it IRAS}\xspace}
\newcommand{\water}{H$_2$O\xspace}
\newcommand{\kms}{ km\,s$^{-1}$\xspace}
\newcommand{\miriad}{{\sc Miriad}\xspace}
\newcommand{\um}{{$\mu{}m$}\xspace}
\newcommand{\degree}{$^{\circ}$\xspace}

\shorttitle{Accurate OH maser positions from the SPLASH pilot region}
\shortauthors{Qiao et al.}

\begin{document}

\title{Accurate OH maser positions from the SPLASH pilot region}

\author{Hai-Hua Qiao\altaffilmark{1, 2, 3, 4}, Andrew J. Walsh\altaffilmark{3}, James A. Green\altaffilmark{5}, Shari L. Breen\altaffilmark{6, 5}, J. R. Dawson\altaffilmark{7, 5}, Simon P. Ellingsen\altaffilmark{8}, Jos\'e F. G\'omez\altaffilmark{9}, Christopher H. Jordan\altaffilmark{3, 10}, Zhi-Qiang Shen\altaffilmark{1, 4}, Vicki Lowe\altaffilmark{11, 5}, Paul A. Jones\altaffilmark{11}}
% , Hiroshi Imai\altaffilmark{6}, James A. Green\altaffilmark{7}, , , , , , Steven J. Gibson\altaffilmark{12}, Maria R. Cunningham\altaffilmark{11}}

\altaffiltext{1}{Shanghai Astronomical Observatory, Chinese Academy of Sciences, 80 Nandan Road, Shanghai, China, 200030; qiaohh@shao.ac.cn} 
\altaffiltext{2}{University of Chinese Academy of Sciences, 19A Yuquanlu, Beijing, China, 100049} 
\altaffiltext{3}{International Centre for Radio Astronomy Research, Curtin University, GPO Box U1987, Perth WA 6845, Australia}
\altaffiltext{4}{Key Laboratory of Radio Astronomy, Chinese Academy of Sciences, China}
\altaffiltext{5}{CSIRO Astronomy and Space Science, Australia Telescope National Facility, PO Box 76, Epping, NSW 2121, Australia}
\altaffiltext{6}{Sydney Institute for Astronomy (SIfA), School of Physics, University of Sydney, NSW 2006, Australia}
\altaffiltext{7}{Department of Physics and Astronomy and MQ Research Centre in Astronomy, Astrophysics and Astrophotonics, Macquarie University, NSW 2109, Australia}
\altaffiltext{8}{School of Physical Sciences, Private Bag 37, University of Tasmania, Hobart 7001, TAS, Australia}
\altaffiltext{9}{Instituto de Astrof\'{\i}sica de Andaluc\'{\i}a, CSIC, Glorieta de la Astronom\'{\i}a s/n, E-18008, Granada, Spain} 
\altaffiltext{10}{ARC Centre of Excellence for All-sky Astrophysics (CAASTRO)}
\altaffiltext{11}{Department of Astrophysics and Optic, School of Physics, University of New South Wales, Sydney, NSW 2052, Australia}

%\altaffiltext{6}{Department of Physics and Astronomy, Graduate School of Science and Engineering, Kagoshima University, 1-21-35 Korimoto, Kagoshima 890-0065, Japan}

%\altaffiltext{12}{Department of Physics and Astronomy, Western Kentucky University, 1906 College Heights Blvd, Bowling Green, KY 42101 U.S.A.}
\begin{abstract}
We report on high spatial resolution observations, using the Australia Telescope Compact Array (ATCA), of ground-state OH masers. These observations were carried out toward 196 pointing centres previously identified in the Southern Parkes Large-Area Survey in Hydroxyl (SPLASH) pilot region, between Galactic longitudes of $334^{\circ}$ and $344^{\circ}$ and Galactic latitudes of $-2^{\circ}$ and $+2^{\circ}$. Supplementing our data with data from the MAGMO (Mapping the Galactic Magnetic field through OH masers) survey, we find maser emission towards 175 of the 196 target fields. We conclude that about half of the 21 non-detections were due to intrinsic variability. Due to the superior sensitivity of the follow-up ATCA observations, and the ability to resolve nearby sources into separate sites, we have identified 215 OH maser sites towards the 175 fields with detections. Among these 215 OH maser sites, 111 are new detections. After comparing the positions of these 215 maser sites to the literature, we identify 122 (57 per cent) sites associated with evolved stars (one of which is a planetary nebula), 64 (30 per cent) with star formation, two sites with supernova remnants and 27 (13 per cent) of unknown origin. The infrared colors of evolved star sites with symmetric maser profiles tend to be redder than those of evolved star sites with asymmetric maser profiles, which may indicate that symmetric sources are generally at an earlier evolutionary stage.
\end{abstract}

\section{Introduction}
\label{introduction}
Ground-state ($^{2}{\Pi}_ {3/2}$, $J = 3/2$) hydroxyl (OH) masers were first discovered towards several Galactic \hii regions (e.g. W3(OH)) by \citet{Wee1965} and \citet{Gu1965}. Since their discovery, extensive work on ground-state OH masers has shown that they are commonly associated with regions of high-mass star formation (HMSF) (\citealt{Are2000}; \citealt{Ede2007}) and these masers are predominantly strong in main-line transitions ($F = 1 \to 1$, $F = 2 \to 2$) (i.e. 1665\,MHz and 1667\,MHz; \citealt{Ree1981}). \citet{Qie2014} compiled a catalog of $\sim$375 ground-state OH maser sites associated with HMSF from the literature, representing an up to date compilation of this category of sources. Ground-state OH masers are also detected towards circumstellar envelopes of evolved giant and supergaint stars, such as Miras and OH/IR stars (\citealt{Nge1979}; \citealt{Sea1997}, \citeyear{Seb1997}, \citeyear{Sea2001}), and are typically dominated by emission from the 1612 MHz satellite line (i.e. $F = 1 \to 2$, 1612\,MHz; \citealt{Ree1981}). The maser emission in the 1720\,MHz satellite-line transition ($F =2 \to 1$), is usually associated with supernova remnants (SNRs) (\citealt{Goe1968}) and traces the interaction between SNRs and surrounding dense molecular clouds. Ground-state OH masers are also found associated with planetary nebulae (PN; \citealt{Use2012}), comets \citep{Gee1998} and the centres of active galaxies \citep{Bae1982}.

Previous work on ground-state OH masers has targeted regions likely to have maser emission, such as Infrared Astronomical Satellite (\IRAS) point sources with infrared (IR) colors indicative of high-mass young stellar objects (YSOs) \citep{Ede2007}, circumstellar envelopes of evolved stars with \water and/or SiO masers \citep{Lee1995} and SNRs \citep{Fre1996}. About 30 years ago, a single-dish blind survey of OH main-line transitions was carried out toward a thin strip of the southern Galactic plane between Galactic longitudes 233\degree through the Galactic Center to 60\degree and Galactic latitudes within $\pm$0.5\degree (\citealt{CH1980}; \citealt{CH1983a}; \citealt{CH1983b}; \citealt{CH1987}). \citet{Ca1998} also carried out a systematic survey close to the Galactic plane with the Australia Telescope Compact Array (ATCA) in the main-line transitions. However, these observations have favoured HMSF regions because only main-line transitions were observed. \citeauthor{Sea1997} (\citeyear{Sea1997}, \citeyear{Seb1997} and \citeyear{Sea2001}) used the ATCA and Very Large Array (VLA) to survey the 1612\,MHz OH line towards a large portion of the Galactic Plane, between Galactic longitudes of 315\degree and 45\degree and Galactic latitudes of $-$3\degree and $+$3\degree. The survey of \citeauthor{Sea1997} observed only the 1612\,MHz OH line and was thus heavily favoured towards detecting evolved star OH masers. Therefore, these previous surveys suffer from biases (e.g. towards HMSF regions or evolved star sites) and thus the full population of each type of astrophysical object mentioned above may not be comprehensively studied.
Although it is clear that ground-state OH masers are detected towards different types of astrophysical objects (HMSF regions, the envelopes of evolved stars, SNRs), the proportion of masers associated with each kind of object remains unknown. Studying the proportion will yield a better understanding of the ground-state OH maser populations and characteristics of their associated astrophysical objects.

In order to reduce biases caused by targeted surveys or surveys only observing specific ground-state OH transitions, the Southern Parkes Large-Area Survey in Hydroxyl (SPLASH) was carried out with simultaneous observations of all four ground-state transitions. SPLASH surveyed 176 square degrees of the southern Galactic plane and Galactic Centre (\citealt{Dae2014}), i.e. between Galactic longitudes of 332\degree and 10\degree and Galactic latitudes of $-$2\degree and $+$2\degree (152 square degrees), plus an extra region around the Galactic Centre, where coverage was extended up to Galactic latitudes $+$6\degree for Galactic longitudes 358\degree to 4\degree (24 square degrees). Unsurprisingly, the survey has detected about 600 sites exhibiting OH maser emission. The observations were conducted with the Australia Telescope National Facility (ATNF) Parkes 64-m telescope and achieved a mean rms (root-mean-square) point-source sensitivity of $\sim$65 mJy in each 0.18\kms channel in maser-optimised cubes, but were limited by the spatial resolution (about 13$\arcmin$). This spatial resolution is insufficient to reliably identify the astrophysical objects associated with the maser emission. Therefore, higher spatial resolution observations are needed, which is the motivation for this work.

The majority of OH masers detected in SPLASH are in the 1612\,MHz satellite-line transition and show a double-horned spectral profile. These masers are evolved star OH masers, which also occasionally exhibit detections in the 1665 and/or 1667\,MHz main-line transitions. Star formation OH masers usually have many spectral features in the main-line transitions. Each spectral feature corresponds to one maser spot on the sky, which is usually considered to arise in a single, well-defined position \citep{Wae2014}. Star formation OH masers preferentially trace a later stage of the HMSF process, compared to methanol masers and water masers (e.g. \citealt{CH1987}; \citealt{Ca1998}; \citealt{Bre2010a}). As mentioned above, OH masers in the 1720\,MHz transition are quite rare and are commonly associated with shocked gas, which can be detected in SNRs (\citealt{Goe1968}), HMSF regions (\citealt{Ca2004}; \citealt{Ede2007}), occasionally asymptotic giant branch (AGB) stars \citep{Seb2001} and PNe \citep{Qie2016}.

%compared to other maser species, e.g. SiO and \water

This paper presents accurate positions of ground-state OH masers, which are obtained from ATCA observations, in the pilot region of SPLASH between Galactic longitudes of 334\degree and 344\degree and Galactic latitudes of $-$2\degree and $+$2\degree.

\section{Observations and Data Reduction}
\label{observation}

Observations were conducted with the ATCA from 2013 October 24 to 2013 October 29 and 2015 January 27, using the 6A configuration. At the frequency of the ground-state OH masers, the 6A array results in a synthesised beam between 4.07$\arcsec\times7.41\arcsec$ and 5.50$\arcsec\times12.53\arcsec$. Observational pointing centres were determined based on the Parkes OH maser detections introduced in \citet{Dae2014}. For masers within one field of view, we chose the pointing centre between them; if the masers were previously observed by the Mapping the Galactic Magnetic field through OH masers (MAGMO; \citealt{Gre2012a}) project, we obtained their accurate positions directly from the MAGMO project (Green, J. private communication). We do not include the MAGMO data in the results presented here and only use them to study the detection statistics. Each region was typically observed with five 6 min snapshot observations for a total on-source integration time of 30 min. 

Primary flux calibration was performed with the standard flux density calibrator PKS B1934$-$638 and bandpass calibration with PKS B0823$-$500. Phase calibrators were chosen to be within 7\degree of each target and the phase calibrators were monitored about every 20 min. 
%No pointing calibration was performed.

The Compact Array Broadband Backend was used to collect data, using the CFB 1M-0.5k mode with 16 zoom bands, each with 2048 channels over 1\,MHz giving 0.5 kHz channel spacings, with all four polarisation products. An analysis of the polarization properties of the masers, including the identification of Zeeman pairs, will be presented in a forthcoming paper. The setup included 3 concatenated zoom bands at each of the 1612 and 1720\,MHz maser transitions, 5 concatenated zoom bands covering the 1665 and 1667\,MHz transitions and 3 concatenated zoom bands at the \hi transition. The channel separation is 0.09\kms. Edge channels were masked out during the data reduction process. Since our observations were targeted towards single-dish maser detections, we know the velocity range of most of the maser emission expected in each source. The velocity coverage over which maser emission was searched for each transition was approximately $-$250 to +210\kms for 1612\,MHz, $-$210 to +140\kms for 1665/7\,MHz and $-$180 to +250\kms for 1720\,MHz in the local standard of rest (LSR) reference frame. The mean rms sensitivity of the ATCA observations is $\sim$65 mJy in each 0.09\kms channel. The sensitivity of the ATCA observations is approximately the same as the original SPLASH survey, but with a factor of two higher velocity resolution.
%The velocity range may miss some extremely high velocity maser spots. 

The \miriad task \textbf{uvlin} was adopted to subtract the continuum emission. The data were then reduced using \miriad standard data reduction routines to produce cleaned and restored data cubes, which were also corrected for the primary beam response (\citealt{Sae1995}). The cubes were then searched (over the area of the primary beam) for maser emission. In order to manage the large volume of data and search masers within a reasonable amount of time, the following method (similar to \citealt{Wae2014}) was used to search for masers:

\begin{enumerate}
\item The full data cube was binned in ten velocity channels (after binning, the channel width becomes 0.9\kms) to maximise the sensitivity to maser spots that have a typical OH maser line-width (based on the Parkes SPLASH spectra; \citealt{Dae2014}).
\item A peak-intensity map (in \miriad a moment $-$2 map) was formed from the whole binned cube. 
\item The peak-intensity map was visually inspected for bright spots, which are the signatures of masers. 
\item Any maser candidate found in the peak-intensity maps was then verified in the full data cube. Our observations resulted in sparse sampling of the uv-plane, which, when combined with very bright maser spots, resulted in side-lobe artefacts in the peak intensity maps. These can be distinguished from real maser spots by visually checking their appearance in the full data cube. A real maser spot will typically be seen as the brightest feature in multiple contiguous channels.
\item For channel ranges excluding the maser emission, binned peak-intensity maps were created to check if there were other weak masers.
\end{enumerate}

The above manual method of searching for and characterising features is at least as accurate as an automated method (\citealt{Wae2012}, \citeyear{Wae2014}). After identifying maser spots, their positions are determined in the following steps. An integrated intensity map (in \miriad a moment 0 map) is created with the unbinned channels over which the maser spot is detected. This integrated intensity map has the highest signal-to-noise ratio for that maser spot and therefore exhibits the most accurate position. If two maser spots are overlapping in velocity, we identify their positions by only integrating over velocities where the maser spot in question dominates the flux density. The \miriad task \textbf{imfit} is used to fit the integrated intensity map and obtain both the position and relative uncertainty in the position of the maser spot. The peak velocity and peak flux density of a maser spot are determined from the spectrum, extracted at the position of the maser spot using \miriad task \textbf{uvspec}. The integrated flux density is also calculated using the spectrum of the maser spot (see Section \ref{images} for more details).
We used a systematic approach to determine the velocity range for each maser spot: in order to improve the S/N of the spectra, we averaged the data by binning five channels to give a velocity resolution of about 0.45\kms. We calculated the rms noise (1-$\sigma$) of the binned data and identify the velocity range as those channels with flux density in excess of 3-$\sigma$. In addition, we excluded any part of the spectrum with noise. Where spectra appear to show two overlapping velocity components with a trough between them, we classify each peak as a separate spot if the difference in flux density between the peak of the weaker line and the lowest point of the trough is greater than the 1-$\sigma$ noise on the data. This method includes most emission from maser spots, but at the same time does not include weak emission from channels such that the noise may dominate in that velocity range. However, this method is not perfect and so the derived velocity ranges and integrated intensities are intended to be used as a guide. This is particularly the case when two maser spots are blended in both velocity and position.

The absolute uncertainty in maser positions depends on the phase noise during the observations, which is related to the distance between the phase calibrator and the target region and the accuracy of the known locations of the phase calibrators, the locations of the antennas, as well as the atmospheric conditions. A typical value of 1$\arcsec$ for the absolute positional uncertainty is adopted. This number is based on previous OH maser surveys with the ATCA by \citet{Ca1998}, who found absolute positional uncertainties of OH masers were better than 1$\arcsec$. The relative positional uncertainty between maser spots detected in the same field of view can be more accurate than the absolute uncertainty. The 1-$\sigma$\ relative positional uncertainty is approximately $\frac{\theta}{2S/N}$ (\citealt{Ree1988}), where $\theta$ is the convolved size of the source (i.e., the beam size in the case of point sources, such as masers). This allows us to compare the distributions of maser spots within a site on smaller scales than the absolute uncertainty, but does not allow us to compare the relative positions of maser spots with other data sets (e.g. IR survey: GLIMPSE, WISE; \citealt{Bee2003}, \citealt{Chu2009}, \citealt{Wre2010}; as described below) to better than 1$\arcsec$.

Given that there are very few detections of 1720\,MHz masers, we have used that band to investigate the continuum emission associated with star formation OH masers. We concatenated 3 zoom bands at this frequency which resulted in a bandwidth of 2\,MHz. The data were reduced using standard techniques for ATCA continuum data, resulting in a typical rms noise of 5 mJy. The results are detailed in Section \ref{starformation}.

\section{Results}
\label{result}

In the SPLASH pilot region, of the 196 pointing centres identified in the Parkes observations of \citeauthor{Dae2014} (\citeyear{Dae2014}), 175 are detected with maser emission from our dataset and the MAGMO dataset. We did not detect any maser emission toward 21 positions, which are discussed in more detail in Section \ref{non-detection}. From our dataset and the MAGMO dataset, we detect 700 maser spots. The strongest flux density is 114 Jy and the weakest is 0.17 Jy.
%(G343.127$-$0.063) vs. (G343.370$-$1.350)

Maser spots have been grouped into maser sites based on their separations. The size of maser sites is discussed in Section \ref{size}. We identify a total of 215 maser sites, 111 of which are newly detected. Among these 215 maser sites, 162 maser sites are from our observations and 53 maser sites are from the MAGMO data. Details of these 53 MAGMO maser sites will be published later by the MAGMO team. We include these 53 MAGMO maser sites for statistical purposes only. Of these 215 maser sites, 154 show maser emission at 1612\,MHz, 61 have maser emission at 1665\,MHz, 57 show maser emission at 1667\,MHz and 9 show maser emission at 1720\,MHz. The 162 maser sites, which are obtained from our observations, are detailed in Table \ref{maintable}. In column 1, we assign a name for each maser spot, based on the Galactic coordinates of their accurate positions and the frequency (1612, 1665, 1667, 1720\,MHz) plus a letter to identify the maser spots within the same maser site. For each frequency, letters are assigned sequentially, based on their peak velocities (from low to high). Columns 2 and 3 show accurate positions (R.A. and Decl.) for each maser spot, column 4 is the peak flux density and column 5 is the integrated flux density. Columns 6, 7 and 8 are peak, minimum and maximum velocities, respectively. Columns 9, 10 and 11 show the uncertainties in position, i.e. uncertainties in minor axis, major axis and position angle. Column 12 shows the astrophysical identification of each maser site, which is discussed in Section \ref{identification}. The final column indicates whether the maser site is a new detection. There are 44 maser sites which only have one maser spot, while the remaining 171 sites have more than one. The richest maser site (G336.075$-$1.084) has 26 maser spots in total at 1612, 1665 and 1667\,MHz and the second richest maser site (G337.404$-$0.402 from MAGMO) exhibits 21 maser spots at 1665 and 1667\,MHz.

\onecolumngrid
\LongTables
\tabletypesize{\tiny}
\tablewidth{\textwidth}
\tablecaption{\textnormal{Details of 162 maser sites, derived from the ATCA observations.}\label{maintable}}
\begin{center}
\begin{deluxetable*}{lccccrrrrrrll}
\tablehead{
\colhead{Name}&\colhead{R.A.}&\colhead{Decl.}&\multicolumn{2}{c}{Flux Density}&\multicolumn{3}{c}{Velocity(\kms)}&\multicolumn{3}{c}{Relative uncertainty}&\colhead{Comments$^{a}$}&\colhead{Ref.$^{b}$}\\
\colhead{}&\colhead{(J2000)}&\colhead{(J2000)}&\colhead{Peak}&\colhead{Integrated}&\colhead{Peak}&\colhead{Min.}&
\colhead{Max.}&\colhead{Minor}&\colhead{Major}&\colhead{Position}&\colhead{}&\colhead{}\\
\colhead{}&\colhead{($^{\rm h\,m\,s}$)}&\colhead{($^{\circ~\prime~\prime\prime}$)}&\colhead{(Jy)}&\colhead{(Jy\kms)}&\colhead{}&\colhead{}
&\colhead{}&\colhead{axis}&\colhead{axis}&\colhead{angle}&\colhead{}&\colhead{}\\
\colhead{}&\colhead{}&\colhead{}&\colhead{}&\colhead{}&\colhead{}&\colhead{}&\colhead{}&\colhead{(arcsec)}&
\colhead{(arcsec)}&\colhead{($^{\circ}$)}&\colhead{}&\colhead{}\\
}
\startdata
G333.942+0.387-1612A&16:21:01.143&-49:26:14.21&0.42&0.53&-72.9&-74.4&-71.7&0.38&0.59&-12.4&U&N\\
\\
G334.097+0.185-1612A&16:22:34.371&-49:28:12.98&0.78&0.87&-25.8&-26.7&-24.0&0.32&0.49&-12.4&ES-SEV&S97\\
G334.097+0.185-1612B&16:22:34.365&-49:28:12.79&0.21&0.19&1.2&1.0&2.3&0.56&0.88&-12.4&ES-SEV&S97\\
G334.097+0.185-1612C&16:22:34.350&-49:28:12.60&0.41&0.58&3.9&2.8&5.1&0.34&0.53&-12.4&ES-SEV&S97\\
\\
G334.132+0.384-1720A&16:21:51.466&-49:18:17.11&0.36&0.31&-45.4&-45.7&-44.4&0.84&1.30&-12.4&U&N\\
\\
\enddata

\tablenotetext{}{Notes: The first column lists the name of the maser spot, which is obtained from the Galactic coordinates of the accurate positions, followed by the frequency and a letter to denote the sequence of maser spots in the spectrum. The second and third columns are the coordinates of the maser spot. The fourth and fifth columns list the peak flux density and integrated flux density. The sixth, seventh and eighth are peak, minimum and maximum velocities, respectively. The ninth, tenth and eleventh columns are minor axis uncertainty, major axis uncertainty and position angle. The twelfth column is the identification of each maser site and the reason for this identification. The thirteenth column identifies new maser sites with ``N'' or gives the reference for previously detected maser sites. Only the 162 maser sites from our observations are shown here.}

\tablenotetext{a}{Sources with unknown assignments are listed as U. The rest of this column is formatted as `Assignment-Reason', where `Assignment' can be SF -- star formation; ES -- evolved star, SN -- supernova remnant or PN -- planetary nebula. `Reason' can be MMB -- class II methanol maser site, based on the Methanol Multibeam data (\citealt{Cae2011}); HOP -- based on HOPS identification \citep{Wae2014}; RMS -- based on RMS identification \citep{Lue2013}; BAI -- \citet{Bae2009}; CAS -- \citet{Ca2004}; CHE -- \citet{Che2012}; GRE -- \citet{Gre2008}; JON -- \citet{Joe1982}; LI1 -- \citet{Tee1991}; LI2 -- \citet{Tee1989}; MCD -- \citet{Mce2012}; SEV -- \citet{Seb1997}; USC -- \citet{Use2014}; VIS -- based on visual check on the GLIMPSE or WISE image together with spectral features.}

\tablenotetext{b}{Sources first detected in this survey are listed as N. The remaining sources are tabulated with references: S97 (\citealt{Seb1997}), L91 (\citealt{Tee1991}), L89 (\citealt{Tee1989}), C98 (\citealt{Ca1998}) and C04 (\citealt{Ca2004}).}
\end{deluxetable*}
\end{center}

\twocolumngrid
\subsection{The maser site images}
\label{images}
\begin{figure*}
\includegraphics[width=0.9\textwidth]{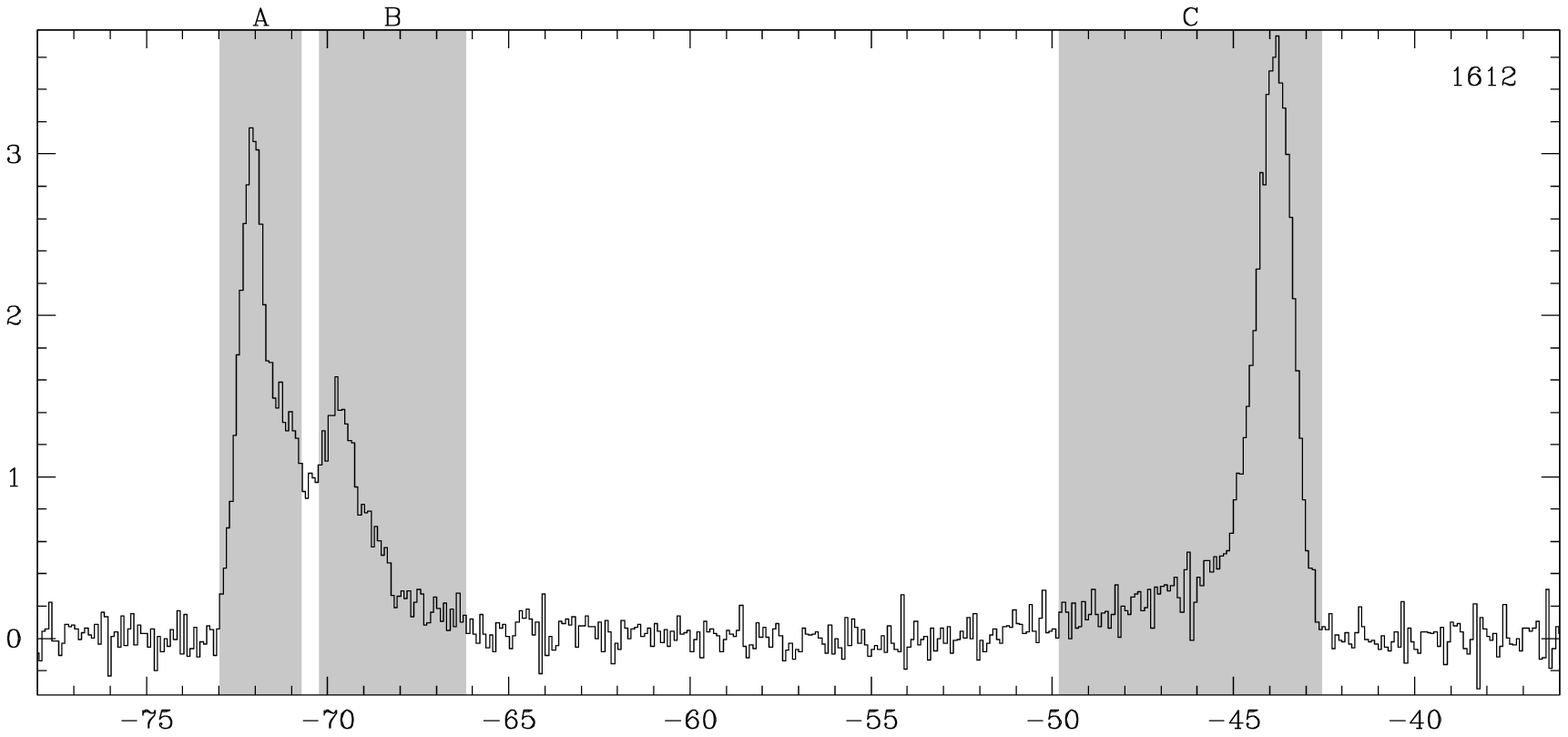}
\includegraphics[width=0.9\textwidth]{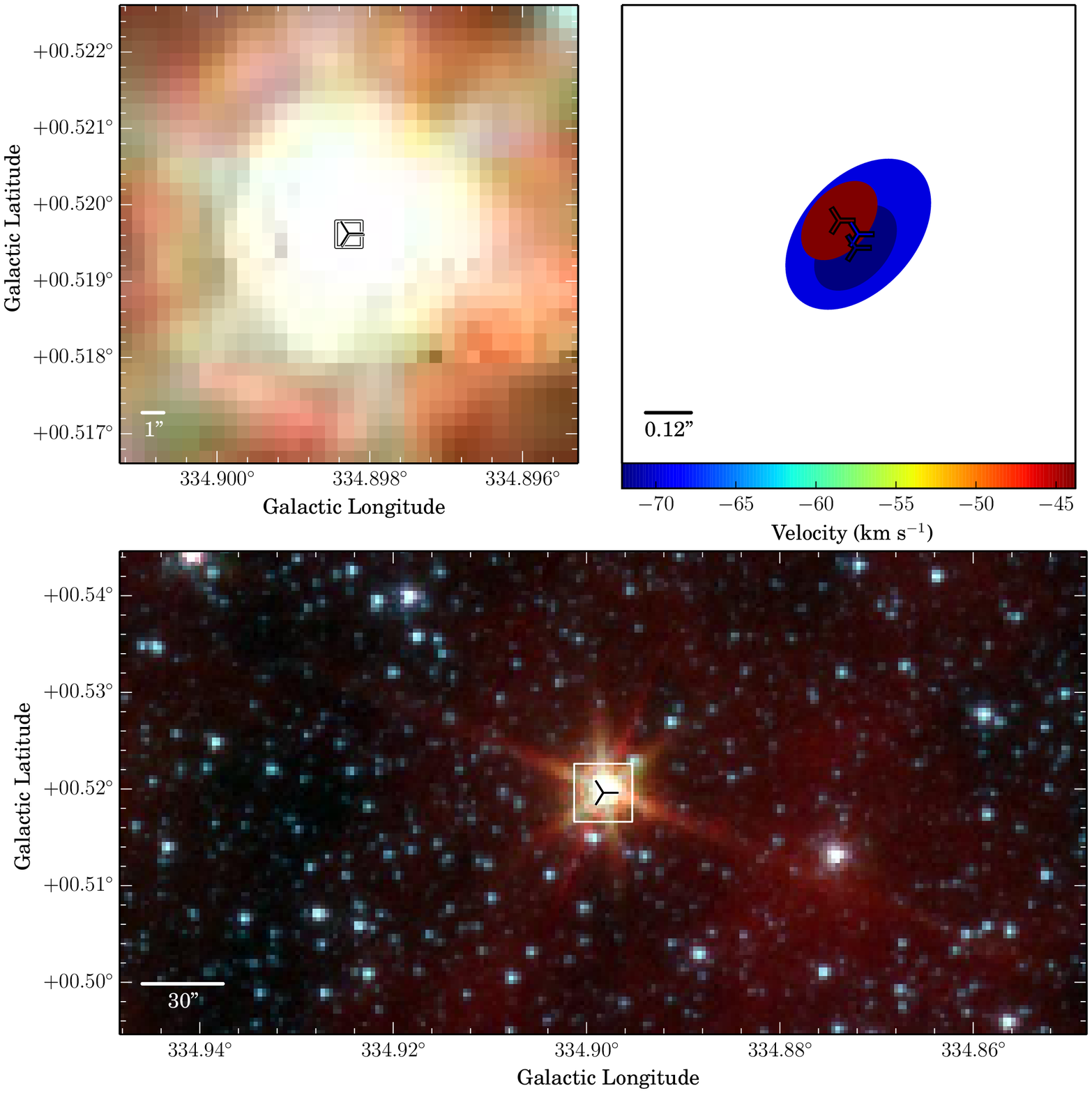}
\caption{G334.898$+$0.520. The upper panel shows the unbinned spectrum, with the radial velocity with respect to the LSR on the x-axis in units \kms and flux density on the y-axis in units of Jy. The flux density is constructed from Stokes I. In the bottom, middle-left and middle-right panels, 1612\,MHz maser spots are presented with 3-pointed stars, 1665\,MHz maser spots with plus symbols, 1667\,MHz maser spots with cross symbols and 1720\,MHz maser spots with triangles. Refer to Section \ref{images} for a full description of the figure. The full 162 figures for this paper are available online.}
\label{figs1}
\end{figure*}

For each maser site, we present the spectrum or spectra, maser spots overlaid on the IR maps and their relative positional error ellipses for each maser spot (e.g. Figure \ref{figs1}, similar to Figure 5 in \citealt{Wae2014}). The upper panel shows the flux density (constructed from Stokes I) plotted against the radial velocity with respect to the LSR for the unbinned spectrum, or spectra (depending on the number of transitions detected), for each maser site. The maser transition is usually labelled in the upper-right corner of the spectrum. The shaded velocity range shows the emission channels for each maser spot, which is labelled with letters at the top. The integrated flux density of each maser spot is the area under the spectral line curve in the shaded channels. Note that the shaded channels were inspected in the full data cube and only include the emission which is easily identified at a single location. Thus, some spectra have features which may appear as peaks and lie outside the shaded regions. In such circumstances, we have identified them as either noise spikes or emission from nearby (but unrelated) strong masers (for individual sources, see details in Section \ref{individual}). Therefore, the shaded channels should be considered as a guide to represent the velocity range over which we believe there is real emission originating from this maser site. Three spectra (Figures \ref{figs_bad1}, \ref{figs_bad2} and \ref{figs_bad3}) have zero values in some velocity ranges due to flagged channels. Only the 162 maser sites from our observations are shown here. The remaining 53 MAGMO maser sites will be published later by the MAGMO team.

The bottom panel is a 6$\arcmin$ $\times$ 3$\arcmin$ GLIMPSE three-color image, which plots band 1 for blue, 2 for green and 4 for red, with band wavelengths of 3.6, 4.5 and 8.0 \um, respectively. Note that for masers in the Galactic latitude range of $|b|>1$\degree, the three-color image is made from WISE data, based on band 1 for blue, 2 for green and 3 for red, with band wavelengths of 3.4, 4.6 and 12.2 \um, respectively. WISE data are used in this Galactic latitude range because the region is outside the GLIMPSE survey area. The image is centred on the maser site. All maser spots that were detected within this field of view are shown. 1612\,MHz maser spots are presented with 3-pointed stars, 1665\,MHz maser spots with plus symbols, 1667\,MHz maser spots with cross symbols and 1720\,MHz maser spots with triangles. Occasionally, there are two or more maser sites in the image, which can be used to see the relative positions of all maser sites in the same field of view. A scale bar of 30$\arcsec$ is shown in the lower-left corner of this panel. A white box located in the centre of the panel shows the field of view in the middle-left panel. 

The middle-left panel presents the zoomed-in region around the maser site. This region is a square 21.6$\arcsec$ on each side. The background is the same three-color image as the bottom panel. The masers at different frequencies are shown as different symbols as described above. This figure only shows maser spots associated with single maser site. Thus, when comparing with the bottom panel, we can identify which maser spots are associated with this maser site. A scale bar of length 1$\arcsec$ is shown in the lower-left corner of this panel, indicating the absolute positional uncertainty of the maser sites.
In the centre of this panel, there is a white box, which shows the region of the zoomed area presented in the middle-right panel. The size of the middle-right panel is chosen such that all masers, together with the full extent of their error ellipses, will fit within the panel. Thus, different maser sites have different white box sizes in the middle-left panel.

The middle-right panel is a zoomed-in region showing the positions and relative error ellipses of all maser spots for this maser site. Each maser spot is represented both with a colored ellipse and a colored symbol, with different symbols for different transitions as described above. The symbols are filled with colors and have black borders in order to be seen easily. The position of the ellipse is the fitted position of that maser spot. The major axis, the minor axis and the position angle of the ellipse represents the relative positional uncertainty of the maser spot, which is obtained with the \miriad task \textbf{imfit} in Section \ref{observation}. At the bottom of this panel, there is a velocity color bar. The color of the ellipses and symbols shows the peak velocity of the maser spot. A scale bar is shown in the lower-left corner of this panel. Note that there is no absolute coordinate presented in this panel because the absolute position uncertainty is only about 1$\arcsec$. The relative offsets of maser spot positions shown in this panel are typically on much smaller scales. 

\section{Discussion}
\label{discussion}

\subsection{The identification of maser sites}
\label{identification}
A blind survey of OH masers in four ground-state transitions over a large proportion of Galactic plane allows us to study their associations and the proportion of different associations in an unbiased way. Even though SPLASH is an untargeted survey, it is still sensitivity limited. This factor may affect the detection statistics of evolved star and star formation masers, but it is still important to classify the masers according to which kind of astrophysical objects they are associated with. We can identify some maser sites with reliable tracers of star formation, such as class II methanol maser sites (\citealt{Cae2011}). The double-horned profile in the 1612\,MHz transition, combined with a bright IR stellar-like source in the GLIMPSE or WISE map as in Figure \ref{figs1}, is a reliable indicator of an evolved star. However, for other maser sites, there may not be very clear evidence about their associations. Therefore, we adopt a variety of methods to identify each maser site and assign them to different categories, i.e. an evolved star site (including PNe), a star formation site, a SNR site or an unknown site. Note that although we expect the identifications to be correct for each maser site, it is possible that a small number of our classifications are incorrect. In Table \ref{maintable}, the last column is our identification for each maser site and reasons for this identification. The process of identifying the associations is in the following steps:

\begin{enumerate}
\item An OH maser site was classified as a star formation site if its position is close to a class II methanol maser site. We compared our OH masers to the Methanol Multibeam survey detections for this region (MMB; \citealt{Cae2011}) in Galactic coordinates. The absolute positional accuracy of the MMB survey is about 0.4$\arcsec$. When there is a 0.001\degree difference in one coordinate (Galactic longitude or Galactic latitude), it is equivalent to an angular separation of 3.6$\arcsec$. We chose 3.6$\arcsec$ as the maximum separation for ease of use. We find OH maser sites and methanol maser sites are either closely associated (the separation is less than 3.6$\arcsec$) or well separated (the separation is larger than 12.9$\arcsec$).
%If there is a 0.001\degree difference in both coordinates (Galactic longitude and Galactic latitude), it is equivalent to an angular separation of 5.1$\arcsec$.
%From our results, we find that there are no methanol maser sites which are at smaller distances (none between 3.6$\arcsec$ and 5.1$\arcsec$) or slightly larger distances (none between 5.1$\arcsec$ and 12.9$\arcsec$) to a corresponding OH maser site.

\item Remaining maser sites were correlated with the H${_2}$O southern Galactic Plane Survey (HOPS; \citealt{Wae2014}) identifications, which classified 22\,GHz water masers carefully and reliably. The absolute positional accuracy of HOPS is about 1\arcsec. The largest distance for an association is chosen to be 3.6$\arcsec$, which is the same value used in MMB data. The comparison result also shows that OH maser sites and water maser sites are either closely associated (the separation is smaller than 3.6$\arcsec$) or well separated (the separation is larger than 12.9$\arcsec$).

\item Remaining maser sites were compared with the Red MSX Source (RMS; \citealt{Lue2013}) catalogue. An association threshold of 9.5$\arcsec$ was adopted, corresponding to the astrometry limits of the RMS survey (\citealt{Lue2013}). The OH maser site was assigned to the RMS identification if the separation was smaller than this value.

\item The remaining sources were identified in the literature with SIMBAD\footnote{http://simbad.u-strasbg.fr}. Some sources with RMS identifications (e.g. PN and SNR) were also searched in SIMBAD to confirm the identification or revise the identification.

\item The associations could also be identified according to their IR properties or IR properties combined with spectral features. The IR images of all maser sites were checked to see if their identifications from the previous steps were correct. We also tried to classify some sources which did not have a clear identification based on steps 1$-$4. When masers have double-horned profiles in the 1612\,MHz transition and are associated with bright IR stellar-like sources in the GLIMPSE or WISE images, we classified them as evolved star OH masers. If masers were associated with extended green objects (EGOs; \citealt{Cye2008}) in the IR image, they were categorised as star formation OH masers, because EGOs trace the energetic shocked gas in the interstellar medium, which is the signature of star formation (e.g. \citealt{Che2009}). With this step, we could identify many evolved star OH masers and check identifications from above steps, e.g. HOPS and RMS, to obtain a correct identification. 

\item Any OH masers which could not be reliably identified using the preceding steps were assigned as unknown sites.
%Any OH masers which were not classified by the previous steps were assigned as unknown sites.

\end{enumerate}

From both our dataset and the MAGMO dataset, we identify 122 OH maser sites (57 per\,cent) associated with the circumstellar envelopes of evolved stars (one of which is a PN), 64 OH maser sites (30 per\,cent) associated with SFRs, two maser sites towards SNRs and 27 unknown maser sites (13 per\,cent). Compared to the literature (e.g. \citealt{Seb1997}; \citealt{Ca1998}), about half of the evolved star sites (60/122) are new detections, 39 per\,cent of the star formation sites (25/64) are new detections and almost all of the unknown maser sites (26/27) are new detections.
The identification result above demonstrates that the majority of OH masers originate from the envelopes of evolved stars. We will discuss the transition overlap of the evolved star category and star formation category separately in more detail in Section \ref{overlap}. The only PN site (G336.644$-$0.695; in the evolved star category) has 1612, 1667 and 1720\,MHz OH masers and shows variability in the 1720\,MHz transition at two epochs (see \citealt{Qie2016} for more details). The two SNR sites (G336.961$-$0.111, \citealt{Fre1996}, from MAGMO; G337.802$-$0.053, \citealt{Ca2004}, from our dataset) show single-peak 1720\,MHz OH masers, which trace the interaction of the SNR and surrounding molecular clouds. The unknown OH maser sites usually have one maser spot at 1612\,MHz and are associated with a bright IR stellar-like source in the GLIMPSE or WISE image (20/27). These OH masers may come from the circumstellar envelopes of evolved stars where only one maser spot at 1612\,MHz can be seen. 

\subsection{Size of maser sites}
\label{size}

\begin{figure}
\includegraphics[width=0.4\textwidth]{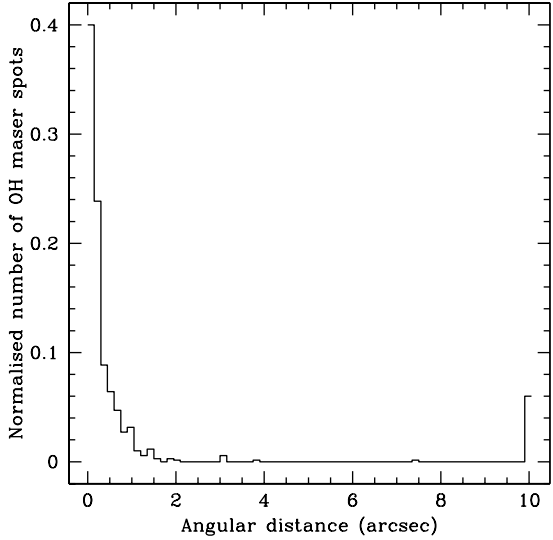}
\caption{Distribution of the angular distance to the nearest neighbour for each maser spot. This distribution quickly falls off with increasing angular distance, indicating that OH maser sites are small and well separated.}
\label{dis_spot}
\end{figure}

\begin{figure}
\includegraphics[width=0.4\textwidth]{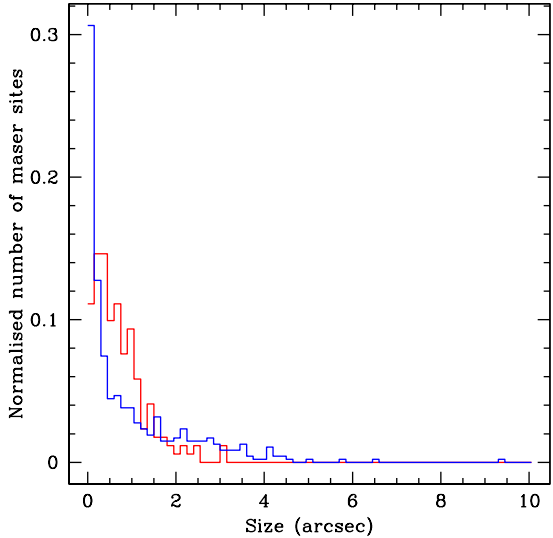}
\caption{Distribution of the sizes of 171 OH maser sites (red) and 470 22\,GHz water maser sites (blue) from HOPS. All these maser sites show more than one maser spot. OH maser sites include evolved star sites, star formation sites and unknown sites. Only eight OH maser sites are equal to or larger than 2\arcsec\ and all of them come from star formation regions. Water maser sites include evolved star sites, star formation sites and unknown sites.}
\label{size_all}
\end{figure}

In order to classify maser spots into maser sites, and thereby determine the upper limit of OH maser site sizes, we have studied the angular separation between each maser spot and its nearest neighbour. Figure \ref{dis_spot} shows that the nearest neighbours are generally located within 2$\arcsec$ of one another. This figure further shows that few spots are separated by values in the range between 2$\arcsec$ and 4$\arcsec$ (a total of five). Given that Figure \ref{dis_spot} plots the separation to the nearest spot (and therefore could be considered a lower limit to site size) we have adopted a site size upper limit of 4$\arcsec$. We note that this is the same upper limit adopted for the water masers detected in the HOPS survey (\citealt{Wae2014}). After establishing an upper limit, the true size of the maser sites can be determined by taking the maximum separation between OH maser spots within that site. In a few cases, we found maser spots spread over more than 4$\arcsec$, but upon inspection were clearly two clusters of spots that were well separated and could therefore be easily classified into two separate sites. Figure \ref{size_all} shows the size of 171 OH maser sites (red) and indicates that all of our OH maser sites are smaller than 3.1$\arcsec$, and only eight sources show spreads of equal to or greater than 2$\arcsec$ (equating to 95 per\,cent of sources). We therefore conclude that the majority of OH masers show extents of less than 2$\arcsec$ (equivalent to a linear size of about 0.05 pc at a distance of 5 kpc) which is consistent with previous investigations (e.g. \citealt{FC1989}). 

For comparison, Figure \ref{size_all} also shows the site sizes of the 470 22\,GHz water maser sites (blue) from HOPS. This shows that the distribution of water maser sizes peaks at a smaller size than OH maser sites, but has a longer tail with more water maser sites showing larger sizes. The result of a Kolmogorov-Smirnov (K-S) test shows that there is a $9\times10^{-4}$ per\,cent chance these two distributions are drawn from the same underlying population. The significant difference in distributions of OH maser sites compared to water maser sites can be accounted for by the fact that very small water maser sites are dominated by evolved star masers (\citealt{Wae2014}), which are found in the inner envelopes compared to OH masers that are associated with the outer envelopes (\citealt{Ree1981}), causing there to be comparatively few very small OH masers. At the other end of the scale, there are more large water maser sites which are dominated by star formation sources (\citealt{Wae2014}) that often trace larger-scale outflows emanating from their powering sources, which may cause the wider range of water maser site sizes. 

\begin{figure}
\includegraphics[width=0.4\textwidth]{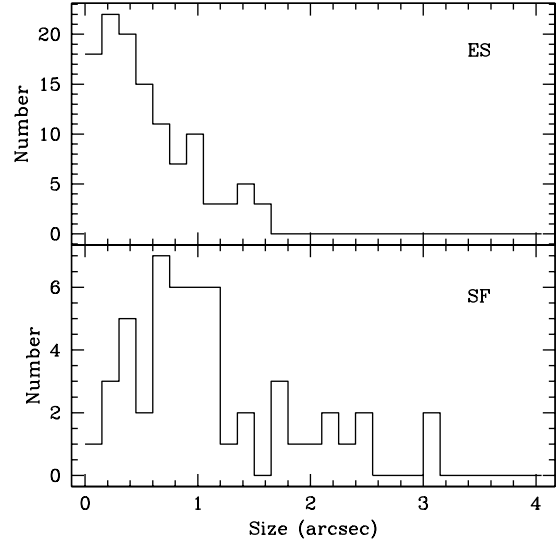}
\caption{Distribution of the sizes of 117 evolved star OH maser sites (top) and 51 star formation OH maser sites (bottom). A K-S test shows that there is a $2\times10^{-5}$ per\,cent chance these two distributions are drawn from the same underlying population.}
\label{size_es_sf}
\end{figure}

Figure \ref{size_es_sf} shows OH maser site sizes for sources associated with evolved stars (117 maser sites) and star formation (51 maser sites), respectively. It can be seen from the plot that evolved star sites are all smaller than 1.6$\arcsec$, while the 12 OH maser sites that show greater extents are all associated with star formation, and, further, that the distribution of the evolved star sites peaks at smaller sizes than the star formation sites. The significance of these differences is confirmed with a K-S test of the two distributions, showing that there is a $2\times10^{-5}$ per\,cent chance they are drawn from the same underlying population. It is perhaps not surprising that evolved star OH maser sites generally tend to be smaller because they are located in a circumstellar shell, typically on size scales of 80 AU (\citealt{Rei2002}), whereas star formation OH maser sites are typically distributed over 3000 AU (\citealt{FC1989}). \citet{Wae2014} also concluded that 22\,GHz water maser sites associated with evolved stars have smaller angular sizes than those sites associated with star formation.

\subsection{Overlap between transitions}
\label{overlap}

\begin{figure}
\includegraphics[width=0.4\textwidth]{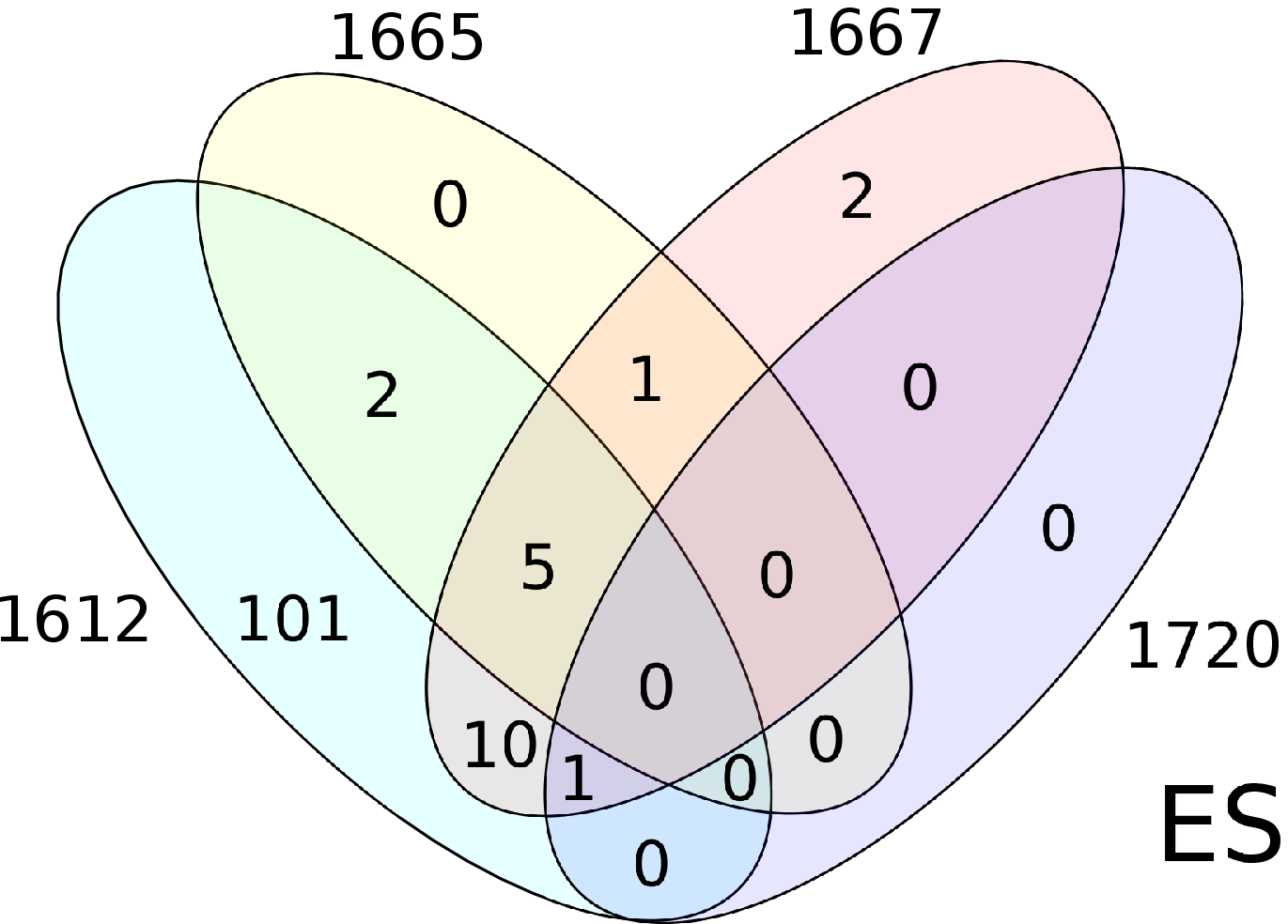}
\includegraphics[width=0.4\textwidth]{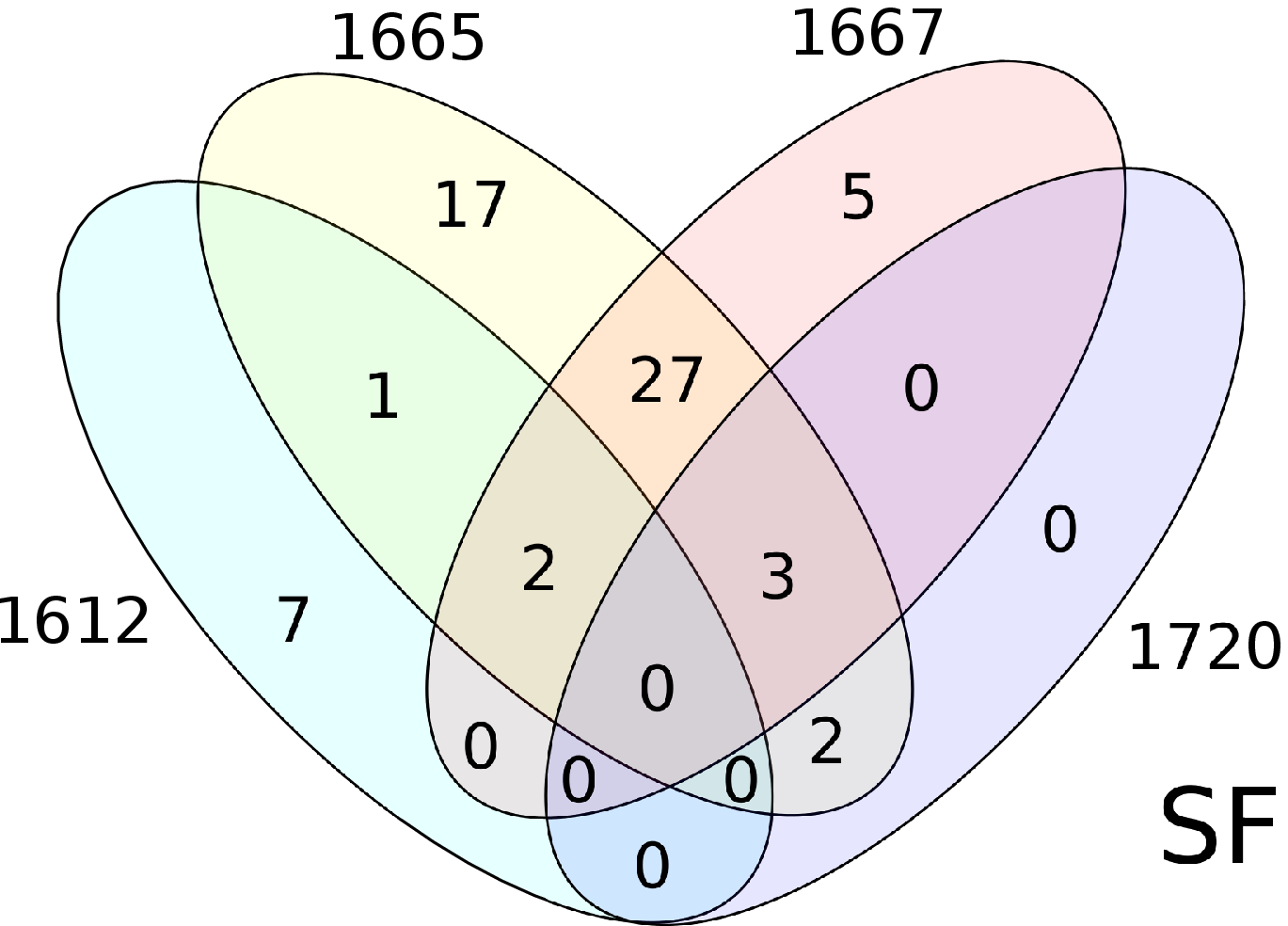}
\caption{Upper panel is Venn diagram showing the overlap of evolved star (ES) OH maser sites at four transitions. Bottom panel is Venn diagram showing the overlap of star formation (SF) OH maser sites at four transitions.}
\label{vennfig}
\end{figure}

Many of the OH maser sites show emission from only a single transition, however, some sites show two or three transitions. The upper panel of Figure \ref{vennfig} shows the transition overlap of evolved star OH maser sites. Of the 122 evolved star maser sites, 101 evolved star sites (83 per\,cent) only exhibit 1612\,MHz emission. Among these 101 evolved star sites, 97 sites have double-horned spectral profiles in their spectra, three have a single peak (G337.064$-$1.173, G338.514$+$1.458 from our dataset and G338.456$-$0.187 from MAGMO) and one has three peaks (G342.004+0.251 from our dataset). Two evolved star sites (G340.043$-$0.092 from our dataset and G336.988$-$0.323 from MAGMO) only show a single peak in the 1667\,MHz transition. One evolved star site only has 1665 and 1667\,MHz masers (G341.083$-$1.084 from our dataset) with two maser spots at each transition. One PN site (G336.644$-$0.695; \citealt{Qie2016}) exhibits the only 1720\,MHz OH maser among evolved stars, and it is accompanied by maser emission at 1612 and 1667\,MHz. 1612 and 1667\,MHz OH masers show the largest overlap: 84 per\,cent of 1667\,MHz OH masers (16/19) have a 1612\,MHz counterpart. Main-line transitions have the second largest overlap with 75 per\,cent of 1665\,MHz OH masers (6/8) associated with 1667\,MHz OH masers.

%Of the 20 remaining evolved star sites, 17 sites show both 1612\,MHz and main-line (1665 and/or 1667\,MHz) masers with two sites showing 1665\,MHz masers, five sites showing both 1665 and 1667\,MHz masers and ten sites showing 1667\,MHz masers. No 1720\,MHz OH masers are associated with evolved stars. 83 per\,cent of 1612\,MHz OH masers are solitary, i.e. without any other OH masing transitions. 

The bottom panel of Figure \ref{vennfig} is Venn diagram illustrating the overlap between the four transitions in star formation regions. 70 per\,cent of 1612\,MHz OH masers (7/10) are solitary, i.e. not associated with other ground-state OH transitions. 33 per\,cent of 1665\,MHz OH masers (17/52) are solitary. Main-line OH masers show the largest overlap: 32 sites exhibit both 1665 and 1667\,MHz OH masers. 62 per\,cent of 1665\,MHz OH masers (32/52) have a 1667\,MHz counterpart, which is higher than the value reported in \citet[53 per\,cent in the SPLASH pilot region]{Ca1998}. \citet{Ca1998} detected 38 star formation OH masers in the SPLASH pilot region and we detect 37 of these. The undetected OH maser (336.984$-$0.183) only showed weak ($\sim$0.4 Jy) emission at 1665\,MHz in \citet{Ca1998} and is not detected here likely due to variability. For the re-detections of \citet{Ca1998}, 72 per\,cent of 1665\,MHz OH masers (26/36) have a 1667\,MHz counterpart. This is partly caused by the broader association radius we used (3.1$\arcsec$ versus 1$\arcsec$ adopted by Caswell, 3.1$\arcsec$ is the largest size of star formation OH maser sites from our results). Moreover, it is possible that the \citet{Ca1998} data have some biases in their 1667\,MHz observations. Since his observations were focused on the 1665\,MHz line, the limited bandwidth of the observations meant that the 1667\,MHz line fell outside the range of the observations in some cases. To combat this, \citet{Ca1998} adopted previous observations where available, meaning that source variability may influence these associations (since the observations of the two lines would necessarily be separated in time). With a larger number of masers from the full SPLASH region (that have been observed simultaneously), we aim to accurately determine the overlap for both evolved star and star formation maser sites.

%A third of the 58 1667\,MHz OH masers (19/58) have a 1612\,MHz OH counterpart and about half of them are also associated with 1665\,MHz OH masers (8/19). Most of these maser sites showing both 1612 and 1667\,MHz emission (15/19) are evolved star sites, which have a double-horned profile in their 1612\,MHz spectra. About one sixth of the 1665\,MHz OH masers (11/62) are associated with 1612\,MHz masers and most of these maser sites also exhibit 1667\,MHz masers (8/11). The majority of these maser sites showing 1612 and 1665\,MHz (7/11) are also from evolved star sites. Five of the nine 1720\,MHz masers detected in this work are associated with 1665\,MHz masers and three of them have both main-line transitions. One 1720\,MHz site also shows 1612 and 1667\,MHz OH masers and is associated with a young PN (G336.644$-$0.695; \citealt{Qie2016}). There are three solitary 1720\,MHz maser sites, two of which originate from SNRs (G336.961$-$0.111, \citealt{Fre1996}, from MAGMO; G337.802$-$0.053, \citealt{Ca2004}). The other one may be associated with a SFR (G334.132+0.384; \citealt{Ca2004}).
%There are seven sites only showing 1612\,MHz maser emission, 16 sites only showing 1665\,MHz maser emission and four sites only showing 1667\,MHz maser emission. 28 star formation sites have both 1665 and 1667\,MHz masers. The remaining eight sites exhibit both main-line transitions and satellite lines.

\subsection{Evolved star sites}
\label{evolvedstar}

For 118 evolved star sites (excluding the PN site) with 1612\,MHz emission, we classified them according to their integrated flux densities of blue- and red-shifted components (I$_{blue}$ and I$_{red}$). Since blue- and red-shifted components come from two sides of the circumstellar envelope, we can estimate the number of photons emitting from each side using the integrated flux density. If the ratio of I$_{blue}$ and I$_{red}$ is between 0.5 and 2, we classify the source as symmetric. If the ratio of I$_{blue}$ and I$_{red}$ is larger than 2 or smaller than 0.5 , we classify them as asymmetric. Three sources (G337.231$+$0.044 from our dataset, G337.078$-$0.187 and G338.456$-$0.187 from MAGMO) are not included because the velocity range of our observations or the MAGMO observations does not cover the expected velocity of either the blue- or red-shifted component. Two sources (G337.064$-$1.173 and G338.514$+$1.458 from our dataset) are classified as extremely asymmetric sources because both of them only have one spectral component in the spectrum.

\begin{figure}
\includegraphics[width=0.4\textwidth]{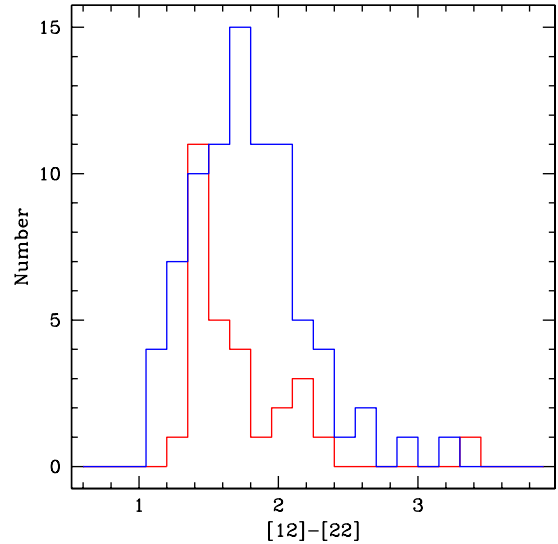}
\caption{WISE [12]-[22] color distributions of 83 symmetric (blue) and 29 asymmetric (red) evolved star sites showing 1612\,MHz transition. A K-S test shows that there is a 10 per\,cent chance that these two distributions are from the same underlying population.}
\label{hist_1222}
\end{figure}

\begin{figure}
\includegraphics[width=0.4\textwidth]{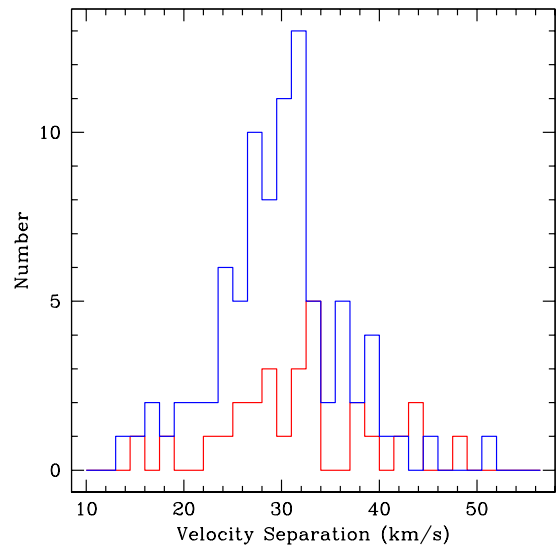}
\caption{The velocity separation $\bigtriangleup V$ of 86 symmetric (blue) and 27 asymmetric (red) evolved star sites showing 1612\,MHz transition. A K-S test shows that there is a 34 per\,cent chance that these two distributions are from the same underlying population.}
\label{Vel_Sep}
\end{figure}

There are 86 (out of 118, 73 per\,cent) symmetric sources and 29 (out of 118, 25 per\,cent) asymmetric sources. We compared the mid-IR properties of these two samples, because emission from circumstellar dust can become the dominant source of emission over radiation from the photosphere for wavelengths longer than 5 \um (\citealt{Ble2006}). 83 (97 per\,cent) symmetric sources have WISE point source counterparts and 29 (100 per\,cent) asymmetric sources have WISE point source counterparts. We find that the [12]-[22] color (12 \um and 22 \um) of these two samples has different distributions, shown in Figure \ref{hist_1222}. This is confirmed with a K-S test showing that there is a 10 per\,cent chance that these two distributions are from the same underlying population. Symmetric sources show redder colors than asymmetric sources, probably indicating that their envelopes are thicker and thus we see colder temperatures. A thicker envelope could indicate that the symmetric sources are at an earlier evolutionary stage than asymmetric ones, and/or that the symmetric sources have more massive precursors and thus, more massive envelopes. Regarding stellar mass, \citet{Smi2003} found OH/IR stars with higher main-sequence masses ($20 < \bigtriangleup V < 40$\kms, $\bigtriangleup V$ is the velocity separation between most blue- and red-shifted maser components in the 1612\,MHz transition) are clearly redder than OH/IR stars with lower main-sequence masses ($\bigtriangleup V \leq 15$\kms). We also studied the velocity separation of symmetric sources and asymmetric sources, shown in Figure \ref{Vel_Sep}. A K-S test shows that there is a 34 per\,cent chance that these two distributions are from the same underlying population, indicating that there is not significant difference in velocity separation (and thus, stellar mass) between symmetric and asymmetric sources. In principle, these results seem to suggest that age is a more determinant factor in the color difference between the symmetric and asymmetric OH sources. However, there could be some biases that need to be considered, such as the particular ratio between I$_{blue}$ and I$_{red}$ we chose to separate these two samples. Moreover, it is possible that the detection of an underlying difference in main-sequence masses requires a sample with a larger size. Further investigation will be carried out with the full SPLASH sample.

\subsection{Star formation sites}
\label{starformation}
\begin{figure}
\includegraphics[width=0.4\textwidth]{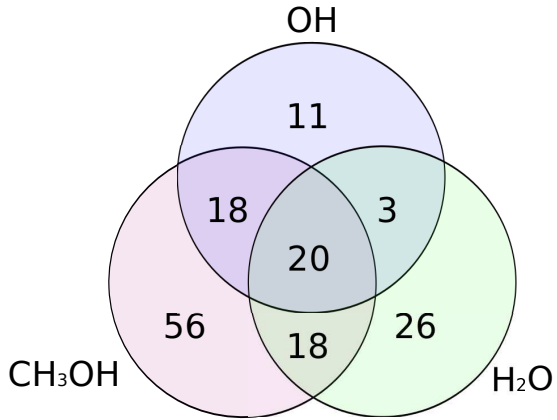}
\caption{Venn diagram showing the overlap of OH, 6.7\,GHz methanol and 22\,GHz water masers from star formation regions, between Galactic longitudes of $334^{\circ}$ and $344^{\circ}$ and Galactic latitudes of $-0.5^{\circ}$ and $+0.5^{\circ}$.}
\label{vennsf}
\end{figure}

We compared star formation OH maser sites in the SPLASH pilot region to 6.7\,GHz methanol maser sites from the MMB survey and 22\,GHz water maser sites from HOPS. We chose the overlap region of these three surveys, i.e. between Galactic longitudes of $334^{\circ}$ and $344^{\circ}$ and Galactic latitudes of $-0.5^{\circ}$ and $+0.5^{\circ}$. There are 52 OH maser sites, 112 methanol maser sites and 67 water maser sites in this region. The association of these three species of masers is shown in Figure \ref{vennsf}. We used the same criteria as described in Section \ref{identification} to identify whether two different maser species are associated. OH maser sites have the largest overlap with methanol maser sites (38/52; 73 per\,cent). As for the OH masers, water masers also have the largest overlap with methanol masers (38/67; 57 per\,cent). Half of methanol maser sites (56/112) are solitary, i.e. not associated with any OH and water masers. The fraction of solitary methanol maser sites is higher than solitary OH maser sites (21 per\,cent) and solitary water maser sites (39 per\,cent). A sensitive 22\,GHz water maser survey towards the MMB masers found that $\sim$48 per\,cent of 6.7\,GHz methanol masers have a 22\,GHz water maser counterpart (\citealt{Tie2014}; \citealt{Tie2016}), which is higher than our result (38/112, 34 per\,cent). Moreover, \citet{Bre2010b} made a targeted search for 22\,GHz water masers toward star formation OH maser sites and concluded about 79 per\,cent of OH masers sites show coincident water maser emission. This value is much higher than our result, which is about 44 per\,cent (23/56). The \citeauthor{Tie2014} and \citeauthor{Bre2010b} water maser surveys have comparable sensitivities with the typical rms noise about 0.1 Jy, whereas HOPS has a typical rms noise of around 1 Jy. Therefore, association results between OH/methanol and water masers will be biased by the lower sensitivity of HOPS. However, it still provides useful insights on the association between OH and 6.7\,GHz methanol masers. We will investigate this overlap with the larger dataset from the full SPLASH survey region.

%20 of these 38 OH maser sites also show water masers. 18 OH maser sites are only associated with methanol masers and three OH maser sites are only associated with 22\,GHz water masers. There are 11 OH maser sites (11/52; 21 per\,cent) which are solitary, i.e. not associated with any methanol and water masers.
%18 of these 38 water maser sites are only associated with methanol masers. There are 26 (26/67; 39 per\,cent) solitary water maser sites.

In the SPLASH pilot region, 64 maser sites are classified as star formation sites. We also checked whether these OH maser sites are associated with continuum sources at 1.7\,GHz. We find six of them (6/64; $\sim$9 per\,cent) are associated with continuum sources. Note that we only used the 2\,MHz bandwidth from 3 concatenated zoom bands for 1720\,MHz to check for continuum sources. A previous study of OH masers in star formation regions shows that $\sim$38 per\,cent of OH maser sites have compact 3\,cm continuum sources (at 8.2\,GHz and 9.2\,GHz ) within 2\arcsec\ (\citealt{FC2000}). For 6.7\,GHz methanol masers, $\sim$20 per\,cent of methanol masers are associated with continuum sources (\citealt{Wae1998}). The low fraction of continuum sources we observe is likely to be due to the low frequency (1.7\,GHz) and limited sensitivity of the continuum observations. Since the turnover frequency of ultracompact \hii regions is typically $> 8-15$\,GHz (\citealt{Kue1994}), ultracompact \hii regions are likely to be optically thick at 1.7\,GHz and the flux density drops off quickly ($S_\nu\propto \nu^{-2}$, where $S_\nu$ is the flux density and $\nu$ is the frequency).

%The rms noise of continuum maps is typically 5 mJy and varies in different maps.

\subsection{Non-detection sources}
\label{non-detection}
Table \ref{nondets} lists the positions at which we did not detect any maser emission in the ATCA observations. From their Parkes spectra, we find all of them are weak (weaker than $\sim$0.4 Jy). Nine of them showed very weak features in the Parkes spectra that we considered as possible masers. But all of these nine could also be consistent with noise. Since we make no detections of these with the ATCA data, we consider the original Parkes identifications as spurious. One of them (G343.95$+$0.35) appears to be diffuse emission, showing line ratios expected for quasi-thermal emission. The remaining eleven positions appear to be real detections with eight showing emission from the 1612\,MHz transition and three exhibiting 1665\,MHz transition. Among these eleven sources, six of them have two peaks and five of them only show one peak in the Parkes spectra. The approximate time difference between the Parkes observations and the current ATCA observations is about 1.5 years. Previous work showed that ground-state OH masers could be variable on short (minutes-hours) and long (days-years) time scales (e.g. \citealt{CJ1991}). Given the higher sensitivity of the ATCA observations, when compared to the previous Parkes observations, we surmise that these masers may be variable and can not be detected because of low flux density during the ATCA observations. 

\tabletypesize{\small}
\begin{table}
\begin{center}
\caption{\textnormal{List of positions which exhibit maser emission in \citet{Dae2014} but were not detected in our observations.}}
\label{nondets}
\begin{tabular}{ccc}
\hline
G334.50$-$0.15(S)&G339.55$+$1.05(S)&G342.55$-$1.85(S)\\
G334.60$+$0.20(V)&G340.40$+$1.15(V)&G342.65$+$0.15(V)\\
G335.95$+$1.80(V)&G341.15$-$0.40(V)&G343.10$+$1.30(V)\\
G337.10$-$0.95(S)&G341.30$+$0.35(V)&G343.25$-$0.05(S)\\
G338.70$-$0.70(V)&G341.55$-$0.70(S)&G343.40$-$0.05(V)\\
G338.70$-$1.05(S)&G341.95$-$1.10(S)&G343.75$+$1.05(V)\\
G339.50$-$0.10(S)&G342.50$-$1.10(V)&G343.95$+$0.35(D)\\

\hline
\end{tabular}
\end{center}

Notes: S -- may be spurious detections; V -- may be variable; D -- diffuse emission, not a maser. The non-detection in half of these regions (11/21) is thought to be due to intrinsic variability, making the masers undetectable in our observations.

\end{table}

\subsection{Comments on individual sources of interest}
\label{individual}

\begin{figure*}
\includegraphics[width=0.9\textwidth]{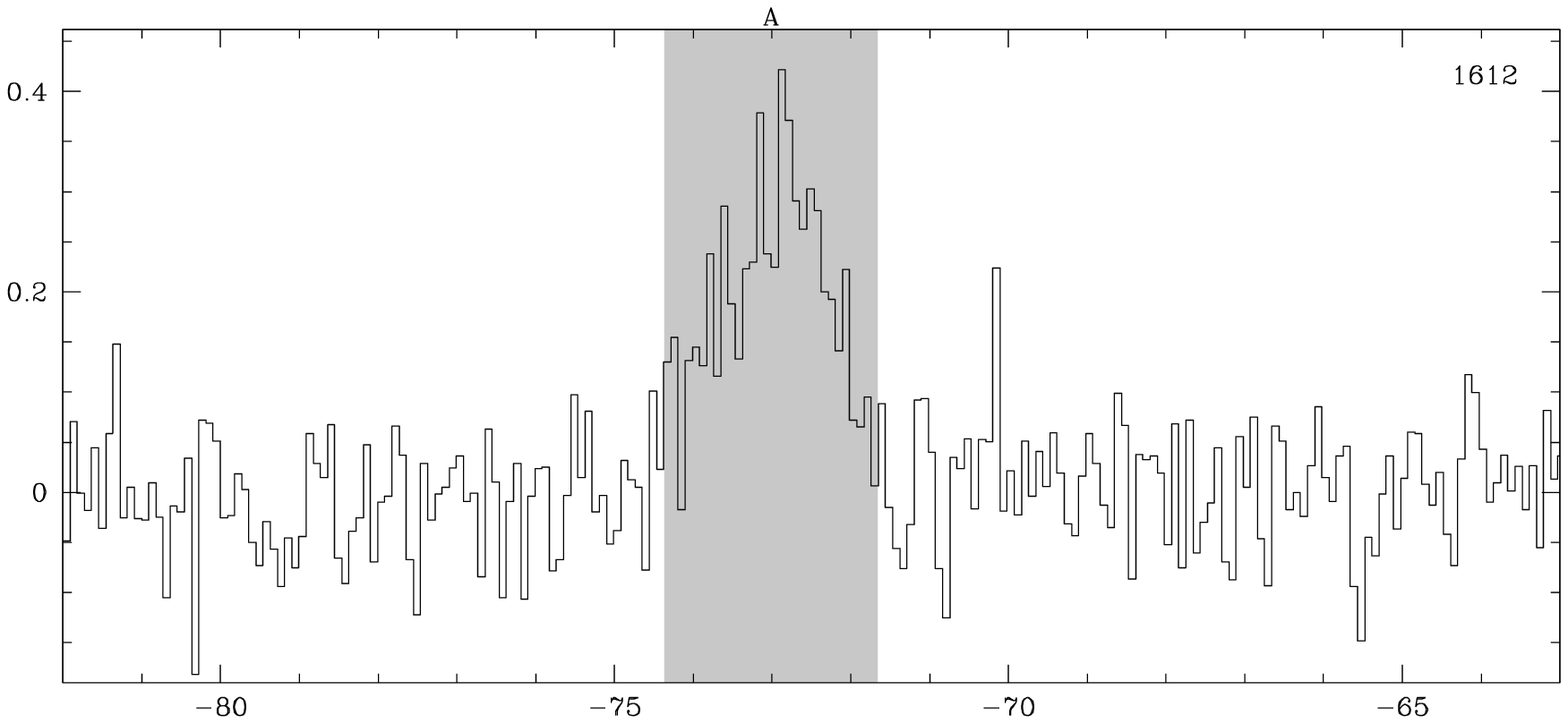}
\includegraphics[width=0.9\textwidth]{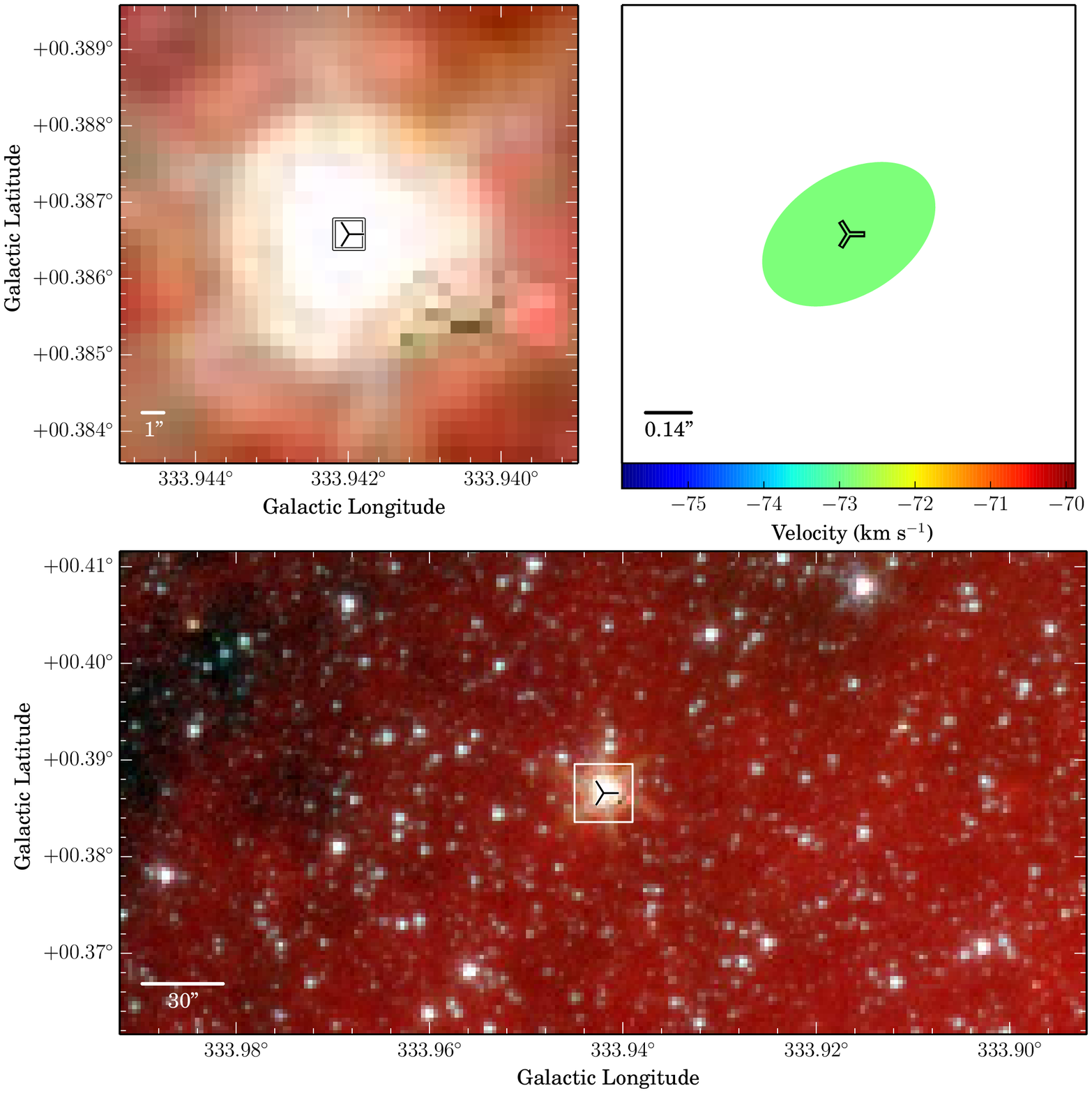}
\caption{G333.942$+$0.387 -- U}
\label{G333.942}
\end{figure*}

\noindent{\bf G333.942+0.387, G334.958$-$0.844,\\ G337.322$-$0.205, G337.877+0.820,\\ G338.902+0.015, G339.871$-$0.670,\\ G342.222+0.041, G342.902$-$0.144,\\ G342.964$-$0.368, G343.191$-$0.279,\\ G343.543+0.345, G343.577$-$0.584\\ and G344.191$-$0.682.} These maser sites only have one OH maser spot in the 1612\,MHz transition. In the GLIMPSE three-color images, they are associated with bright IR stellar-like sources, as shown in Figure \ref{G333.942}. We studied their IR properties by finding their counterparts in the GLIMPSE point source catalogue. For the sources with IR point source counterparts, we found that the magnitude of the 4.5 \um band is brighter than 7.8, which indicates that these sources are ``obscured'' AGB star candidates with very high mass-loss rates according to \citet{Roe2008}. However, in the absence of other confirmation of their nature, we identify them as ``unknown sites'', although they very possibly originate from the circumstellar envelope of evolved stars. 
%(MAGMO, G337.632$-$0.341, G338.199$-$0.154 and G341.841$-$0.049 have similar properties)

\noindent{\bf G334.132+0.384.} This maser site only has one 1720\,MHz maser spot. There is no clear identification for this position in the literature, thus it is identified as an unknown site. From the GLIMPSE three-color image, we find it is located in an extended emission region. It possibly belongs to the variety noted by \citet{Ca2004}, i.e. associated with star formation without main-line emission. This source is resolved on 6\,km baselines, but clearly detected on shorter baselines. It was detected with SPLASH Parkes observations but only in the 1720\,MHz line. We did not detect any emission or absorption towards the other three lines with the ATCA. Thus, we concluded it is likely to be a maser detection.

\noindent{\bf G334.577+1.958, G337.782+1.394,\\ G338.513+1.503 and G340.091+1.458.} These maser sites only have one maser spot at 1612\,MHz. There are no clear identifications for them in the literature, thus their associations are unknown. In the WISE three-color images, they are located in the same position as bright IR stellar-like objects.

\noindent{\bf G335.136+0.196.} This unknown maser site has two 1612\,MHz maser spots. There is no clear identification for this position in the literature. In the GLIMPSE three-color image, it is associated with a bright stellar-like source.

\begin{figure*}
\includegraphics[width=0.9\textwidth]{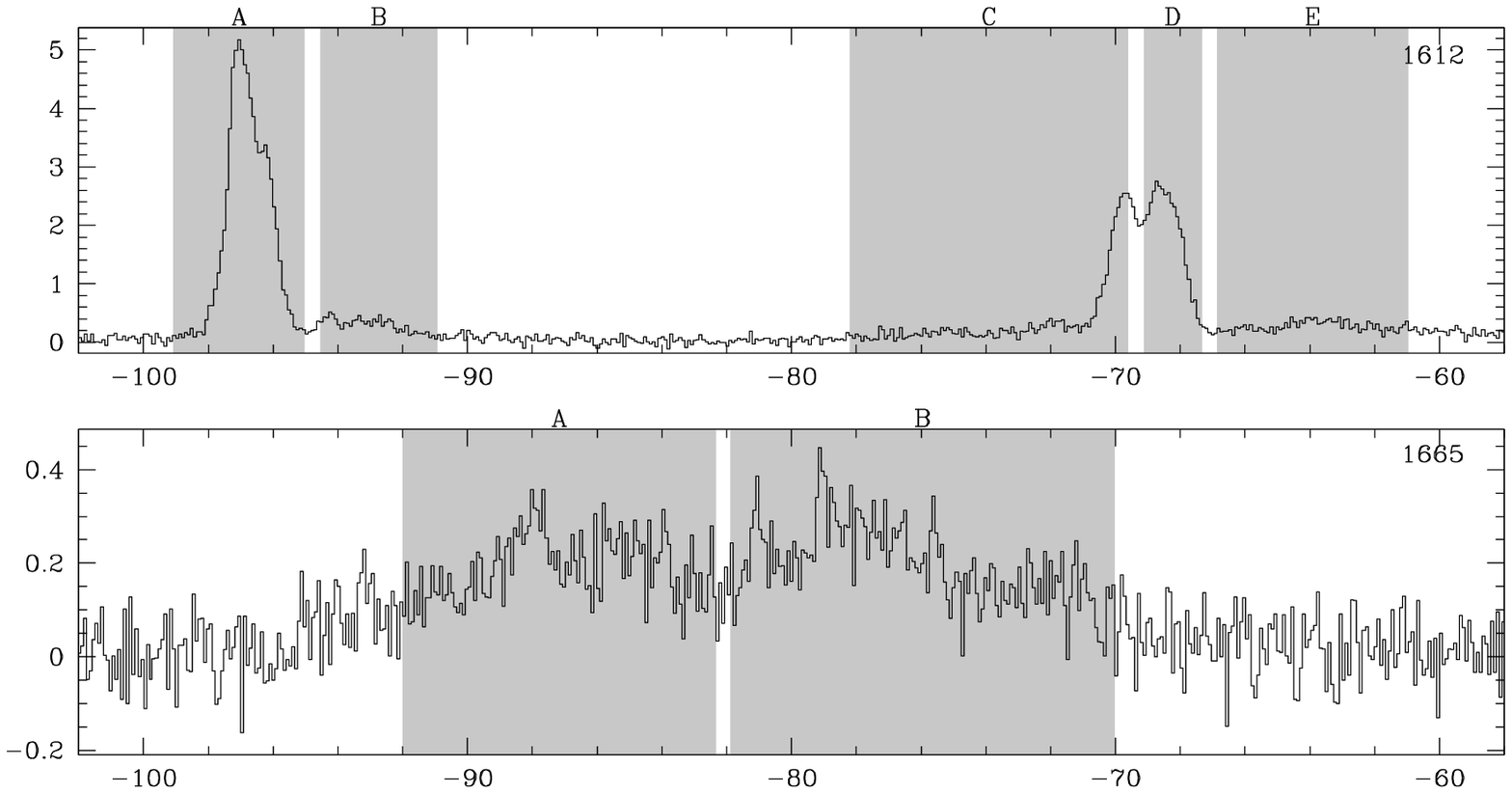}
\includegraphics[width=0.9\textwidth]{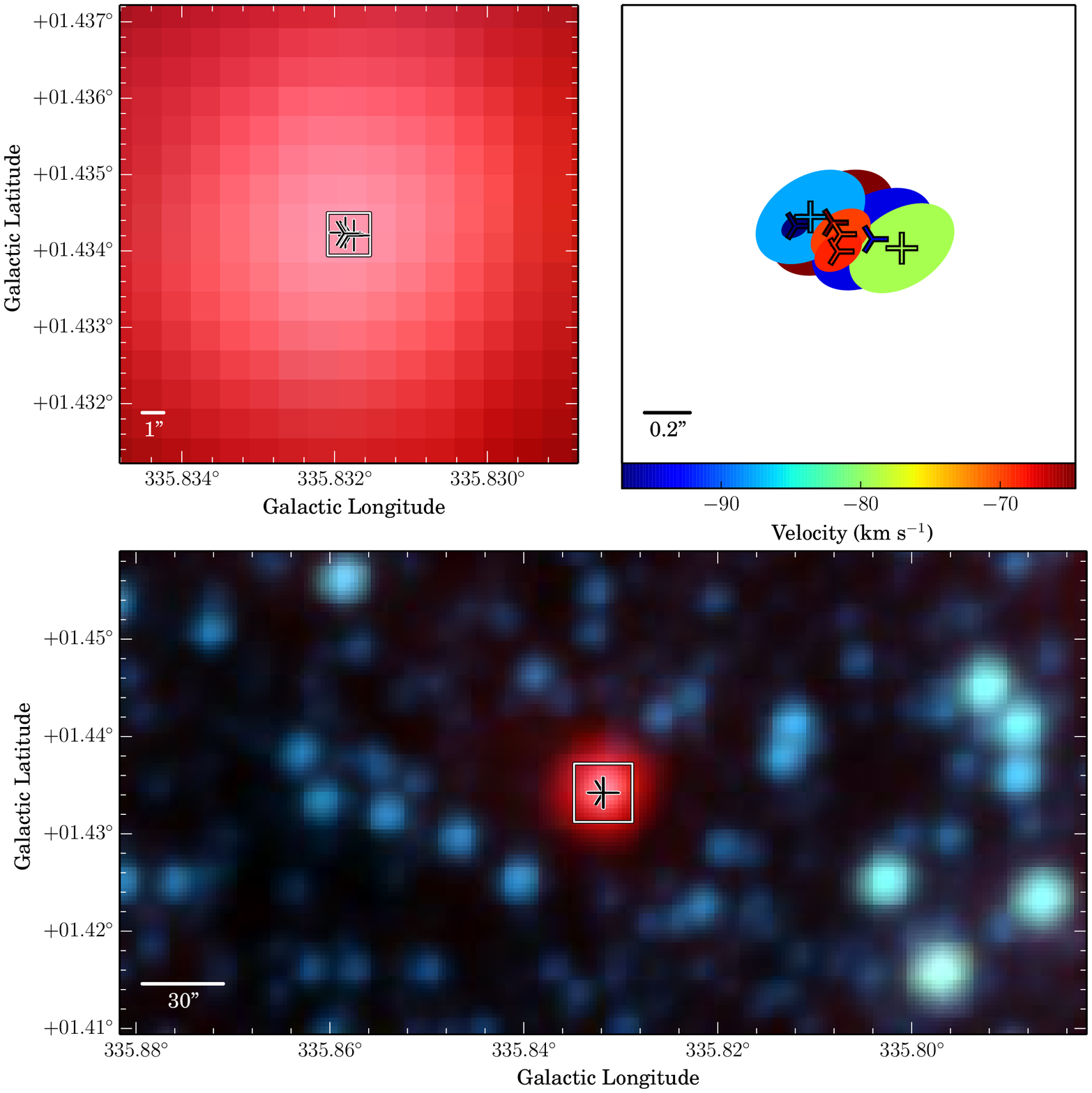}
\caption{G335.832$+$1.434 -- ES}
\label{G335.832}
\end{figure*}

\noindent{\bf G335.832+1.434.} This maser site is an OH/IR star site, which belongs to the evolved star category. Figure \ref{G335.832} shows the line-widths of the 1665\,MHz maser spots are quite broad: about 10\kms for each spot. But since this maser is detected on even the longest baselines and is not detected in the other lines, it is unlikely to be diffuse. In the WISE three-color image, it is associated with a very red star (bright at 12 \um).

\begin{figure*}
\includegraphics[width=0.9\textwidth]{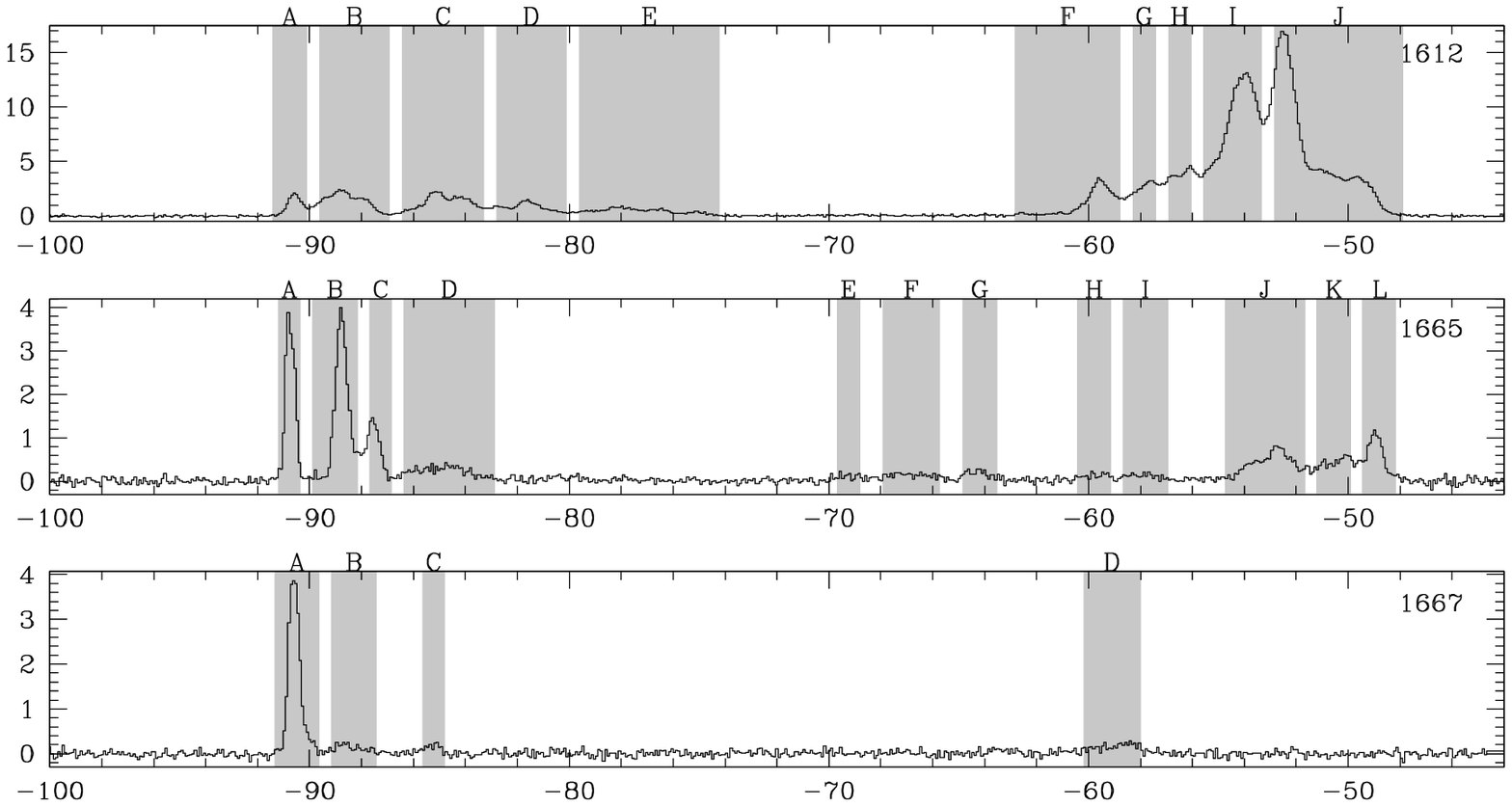}
\includegraphics[width=0.9\textwidth]{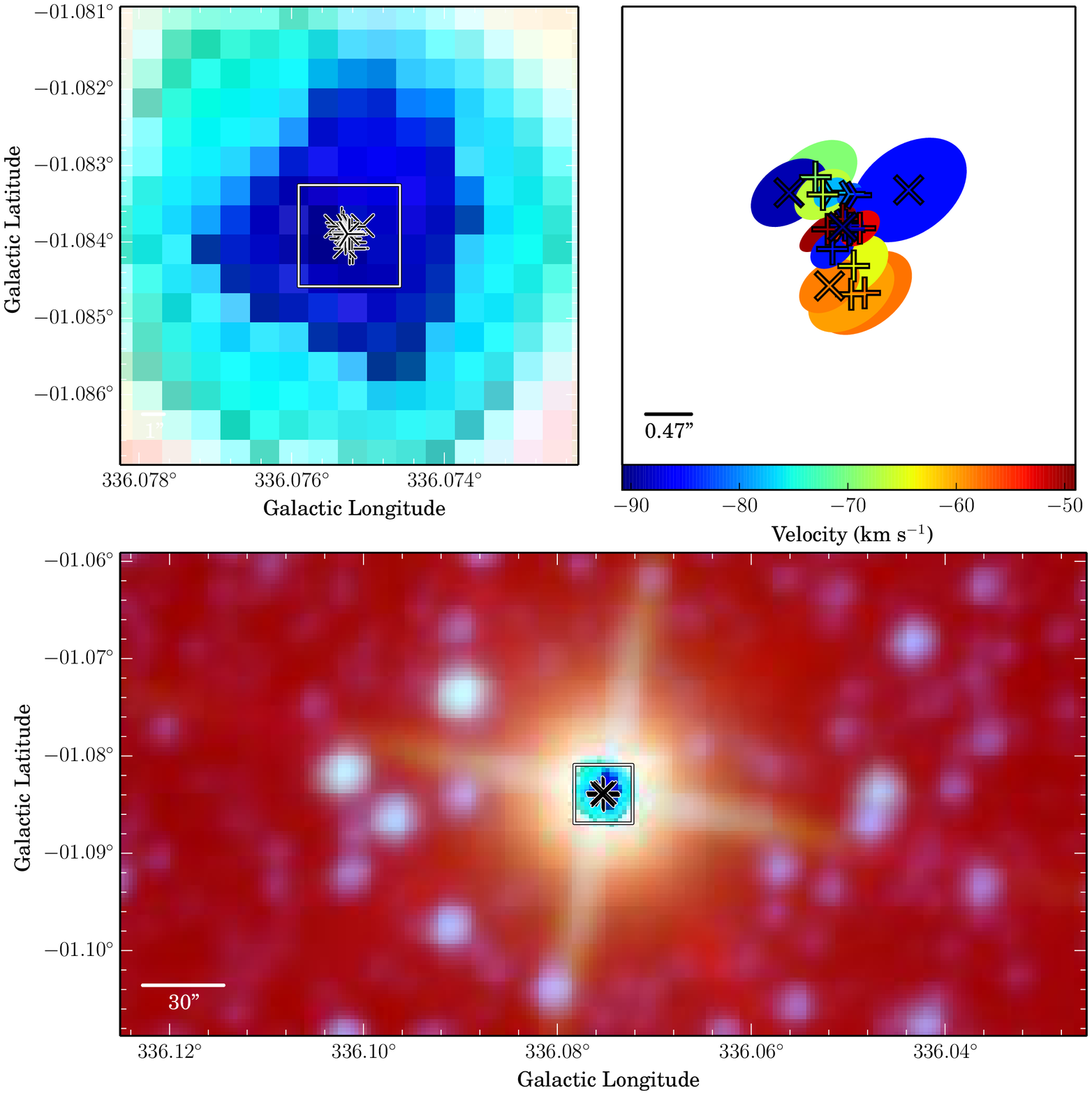}
\caption{G336.075$-$1.084 -- ES}
\label{G336.075}
\end{figure*}

\noindent{\bf G336.075$-$1.084.} This maser site is an OH/IR star site with 26 maser spots at 1612, 1665 and 1667\,MHz, shown in Figure \ref{G336.075}. This is the richest maser site in the SPLASH pilot region. The 1612\,MHz maser spots are stronger in the velocity range $-$62\kms to $-$48\kms, while 1665 and 1667\,MHz maser spots are stronger at the blue-shifted velocity range of $-$92\kms to $-$88\kms. This result indicates that 1665 and 1667\,MHz masers favour a similar physical environment. The GLIMPSE three-color image appears saturated.

\noindent{\bf G336.098$-$0.878.} The unassociated emission in the 1612\,MHz spectrum is from a nearby source, G336.075$-$1.084. %in Figure \ref{figG336.075}. 

\begin{figure*}
\includegraphics[width=0.9\textwidth]{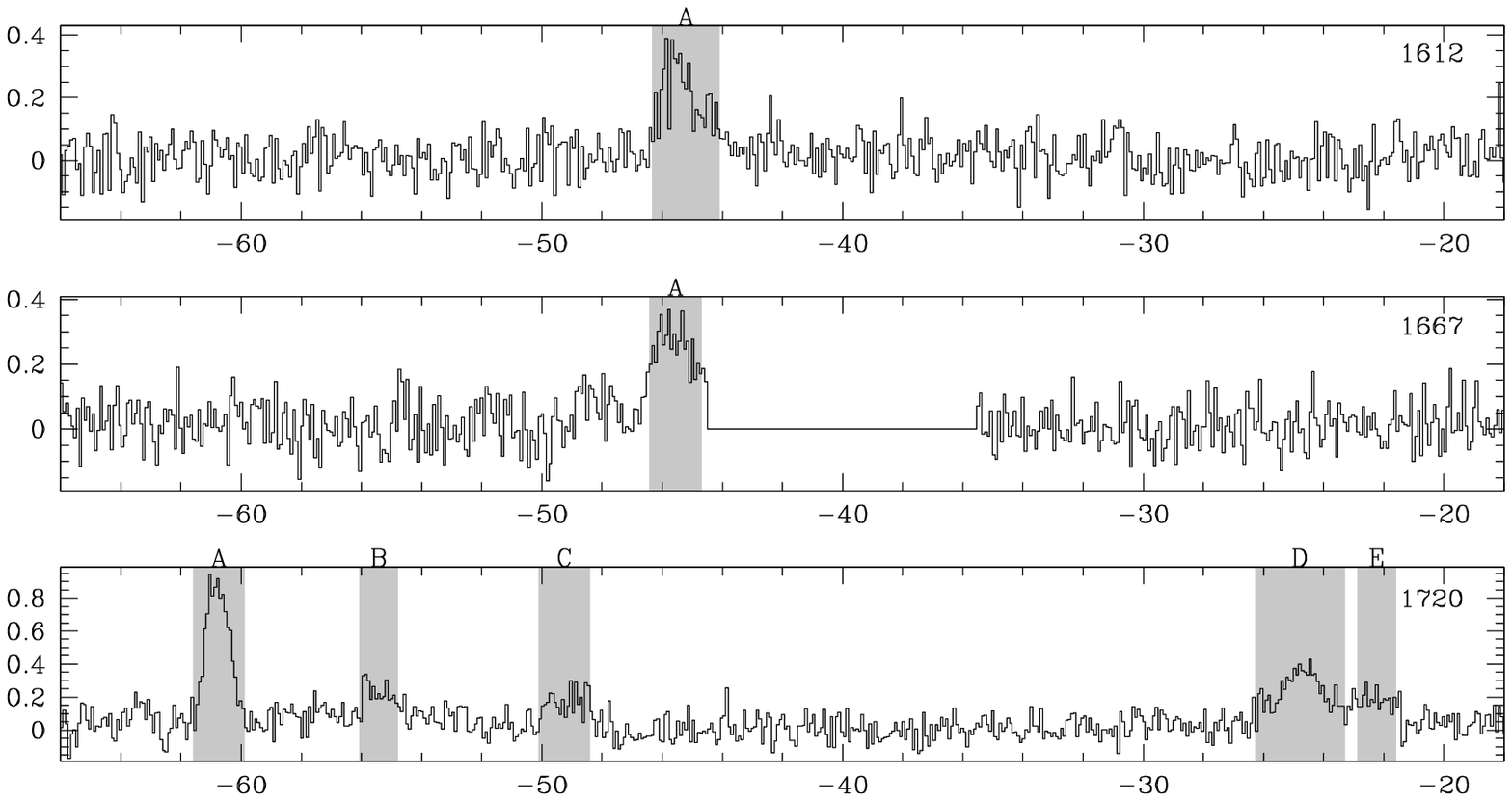}
\includegraphics[width=0.9\textwidth]{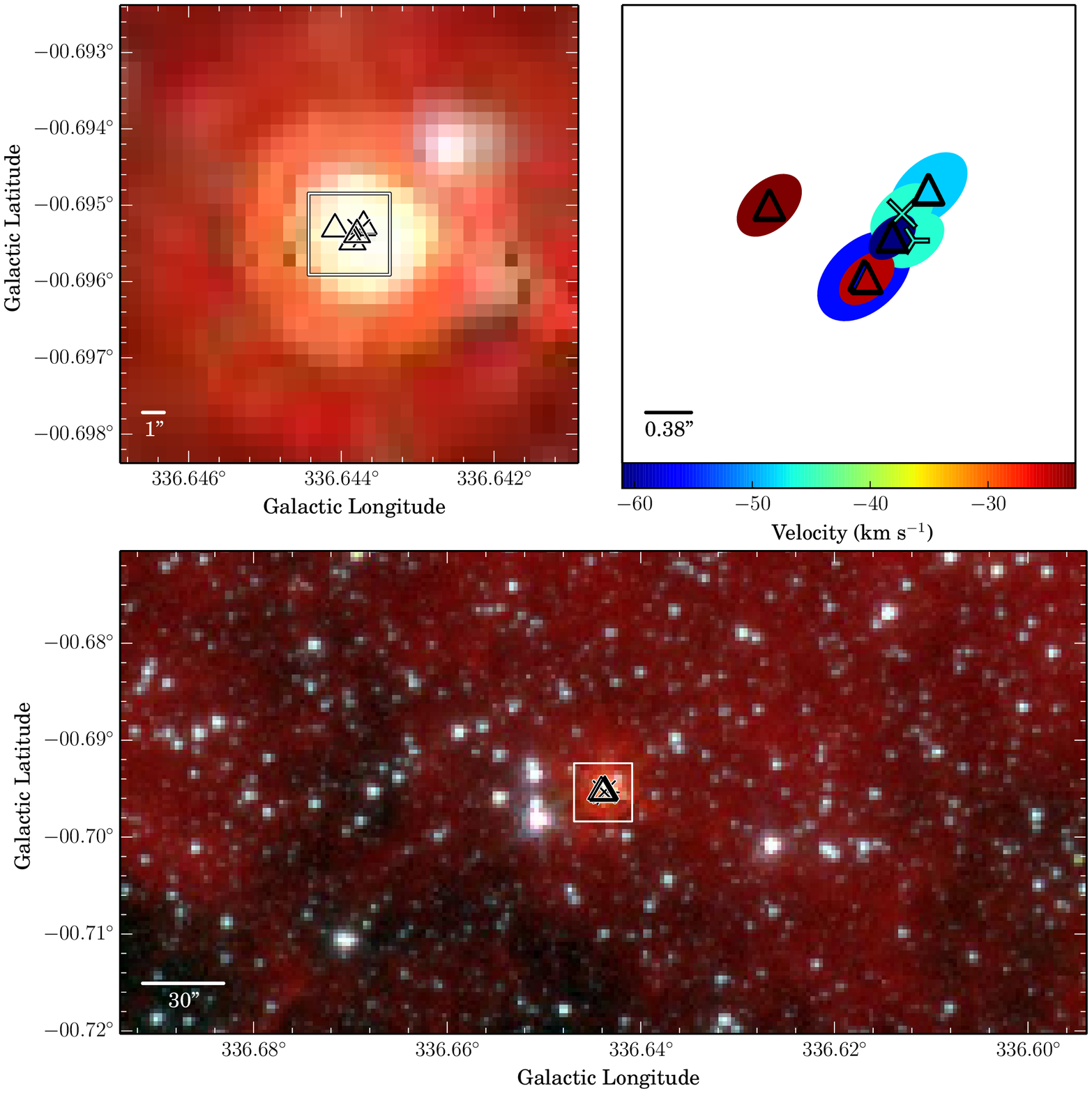}
\caption{G336.644$-$0.695 -- PN}
\label{G336.644}
\end{figure*}

\noindent{\bf G336.644$-$0.695.} This maser site is a PN, which also has 22\,GHz water masers at the velocity of $\sim-$41.8\kms and $\sim-$43.9\kms \citep{Use2014} and radio continuum emission \citep{Vae1993}. As shown in Figure \ref{G336.644}, we detected one 1612\,MHz OH maser spot at $-$45\kms, one 1667\,MHz OH maser spot at $-$45\kms and five 1720\,MHz maser spots, which have the double-horned profile with the central velocity $\sim-$43\kms. This is only the second known PN showing 1720\,MHz OH masers after K 3$-$35 and the only evolved stellar object with 1720\,MHz OH masers as the strongest transition, compared to the other three transitions of ground-state OH. We also observed it with the ATCA about one and half years later and found variability in the 1720\,MHz transition. The magnetic fields are also measured from the Zeeman splitting of the 1720\,MHz maser spots, which suggests that the 1720\,MHz masers are formed in a magnetised environment. Further details of this source can be found in \citet{Qie2016}.

\begin{figure*}
\includegraphics[width=0.9\textwidth]{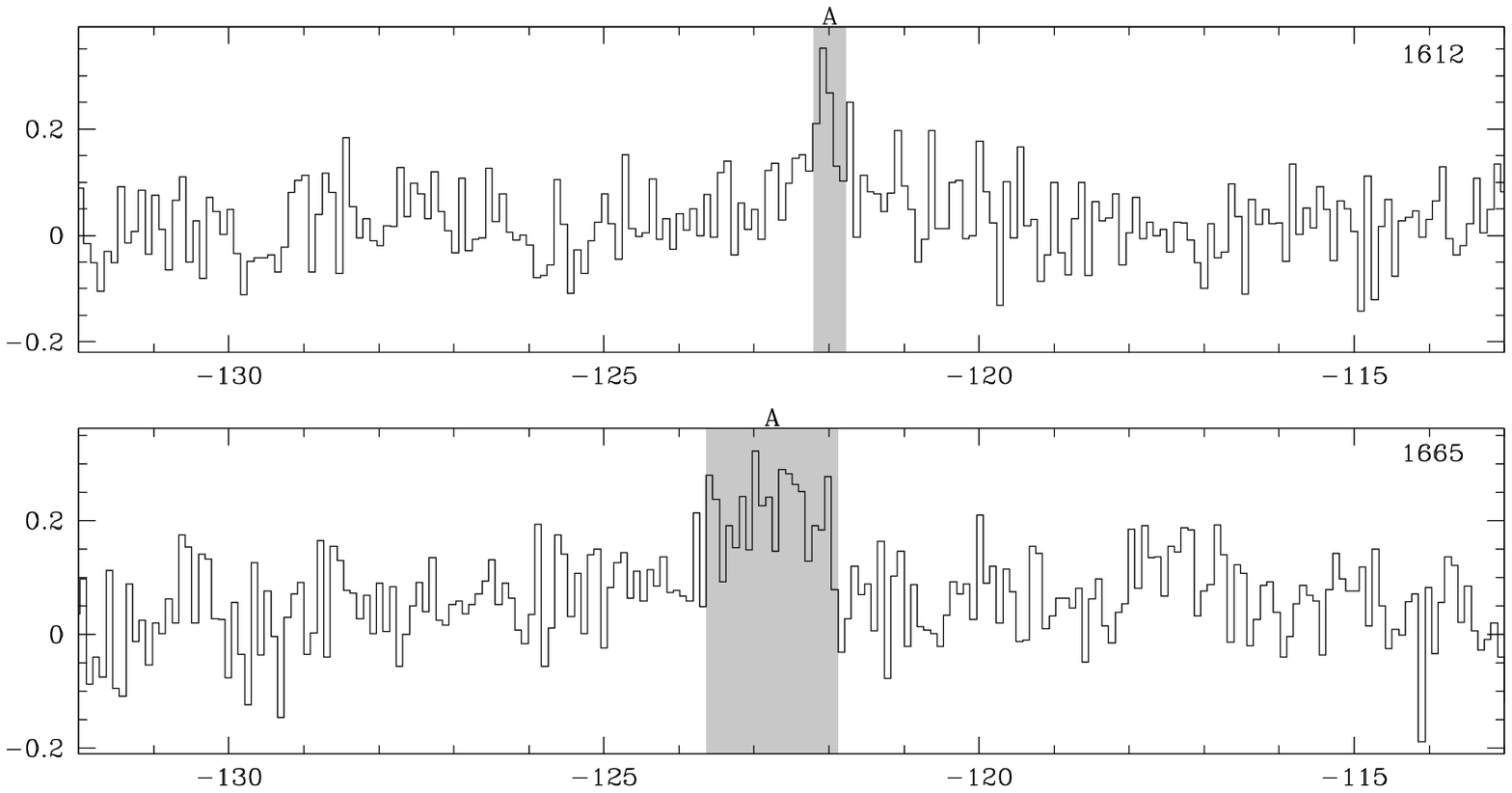}
\includegraphics[width=0.9\textwidth]{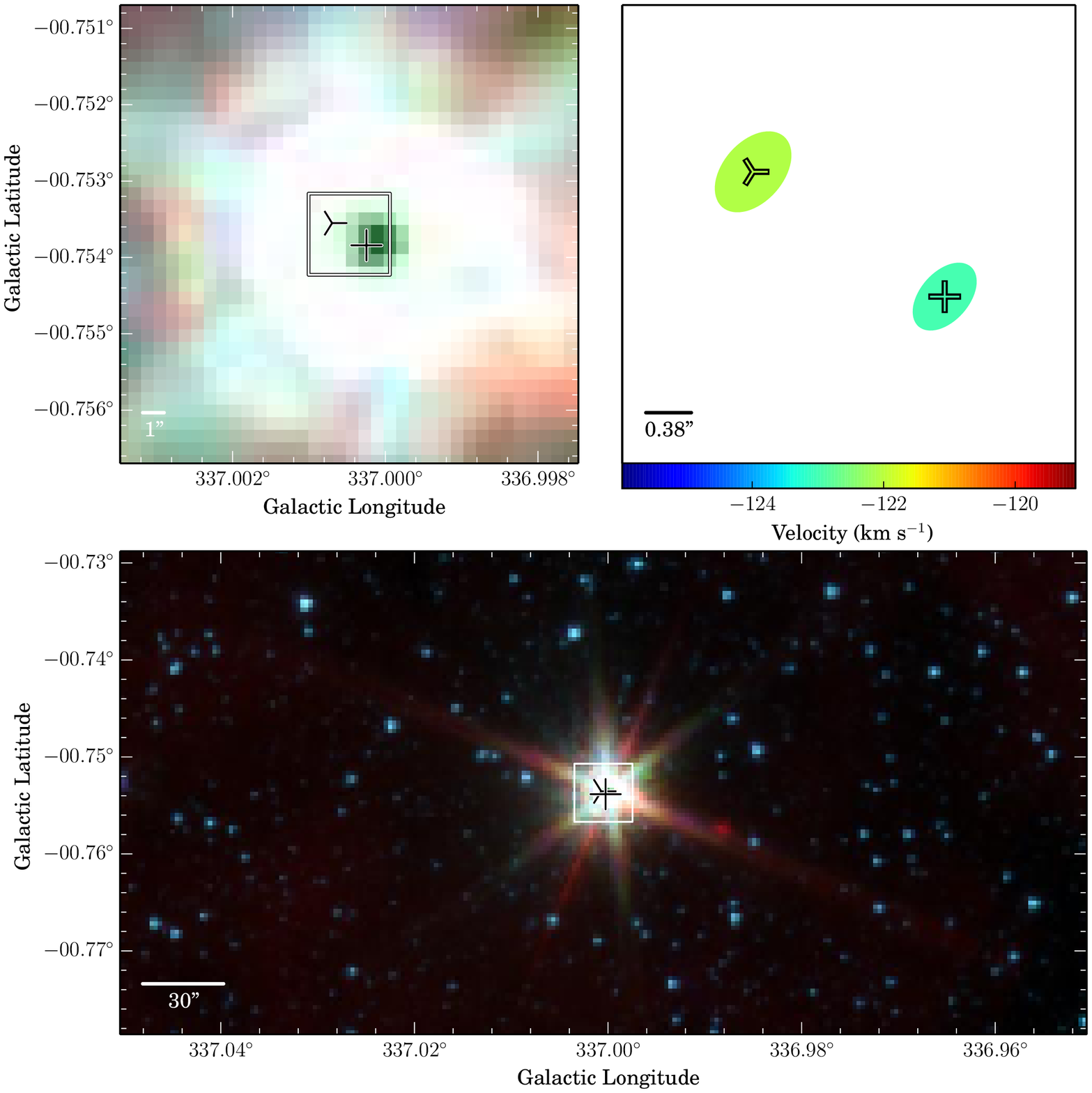}
\caption{G337.001$-$0.754 -- U}
\label{G337.001}
\end{figure*}

\noindent{\bf G337.001$-$0.754.} This maser site is classified as an unknown site. As shown in Figure \ref{G337.001}, it has two maser spots at velocity of $\sim-$122\kms in the 1612 and 1665\,MHz transitions. In the GLIMPSE three-color image, these two masers are associated with a very bright IR stellar-like source, thus it may originate from an evolved star site.

\noindent{\bf G337.064$-$1.173.} This maser site is an evolved star site, which appears to have evolved into the fast wind stage \citep{Bae2009}. This maser has been observed as a single spectral feature by \citet{Seb1997}, which is consistent with our observations. In the WISE three-color image, the maser site is bright at 12 \um (very red).

\noindent{\bf G337.241+0.146.} The unassociated emission in the 1612\,MHz spectrum is from a nearby source, G337.356$-$0.137.% in Figure \ref{figG337.356}

\noindent{\bf G337.802$-$0.053.} This maser site is classified as a SNR site \citep{Ca2004}. It contains one 1720\,MHz OH maser spot, which traces the interaction between the SNR with surrounding molecular clouds. There are no obvious features in the GLIMPSE three-color image.

\begin{figure*}
\includegraphics[width=0.9\textwidth]{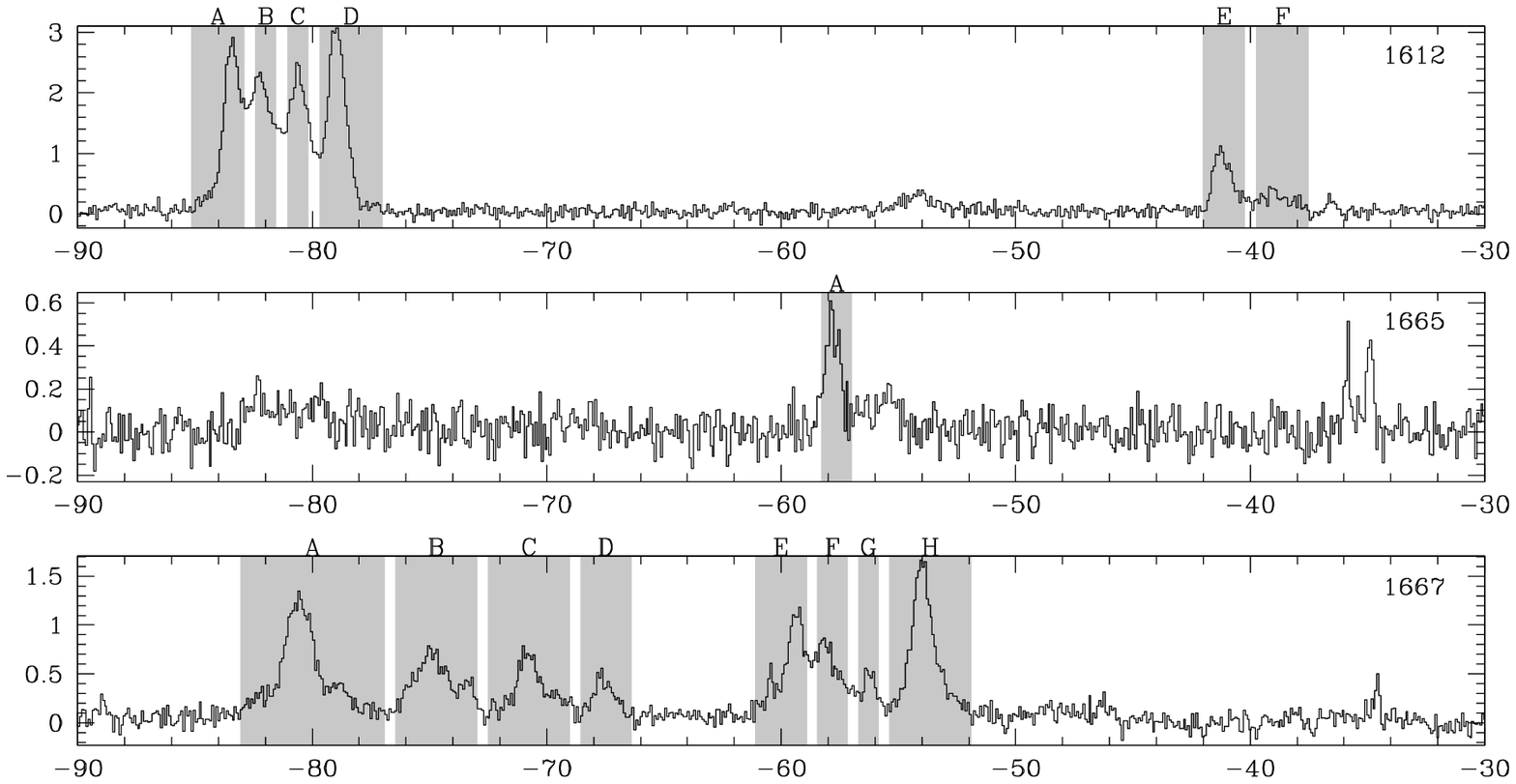}
\includegraphics[width=0.9\textwidth]{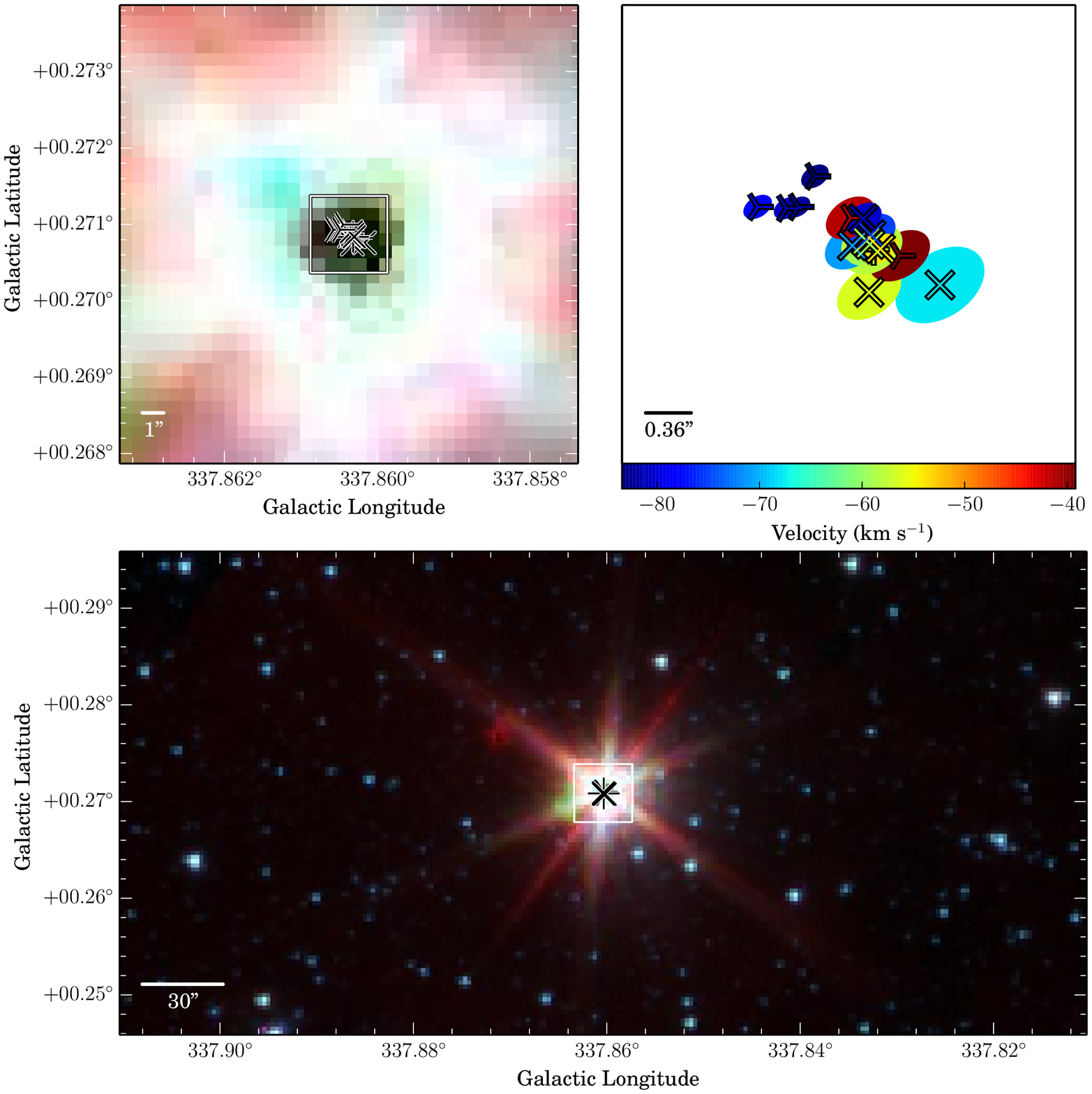}
\caption{G337.860$+$0.271 -- ES}
\label{G337.860}
\end{figure*}

\noindent{\bf G337.860+0.271.} This maser site is an evolved star site, exhibiting 15 maser spots at 1612, 1665 and 1667\,MHz transitions, shown in Figure \ref{G337.860}. The velocity of the 1665\,MHz maser spot is close to the central velocity of the double-horned profile in the 1612\,MHz transition, while the 1667\,MHz maser spots spread a broad velocity range across the 1612\,MHz velocity range. In the GLIMPSE three-color image, these masers are associated with a very bright IR stellar-like source. The unassociated emission in the 1612\,MHz spectrum is from G337.719+0.350 and the unassociated emission in the 1665 and 1667\,MHz spectra is from G337.997+0.136 (the MAGMO dataset).

\begin{figure*}
\includegraphics[width=0.9\textwidth]{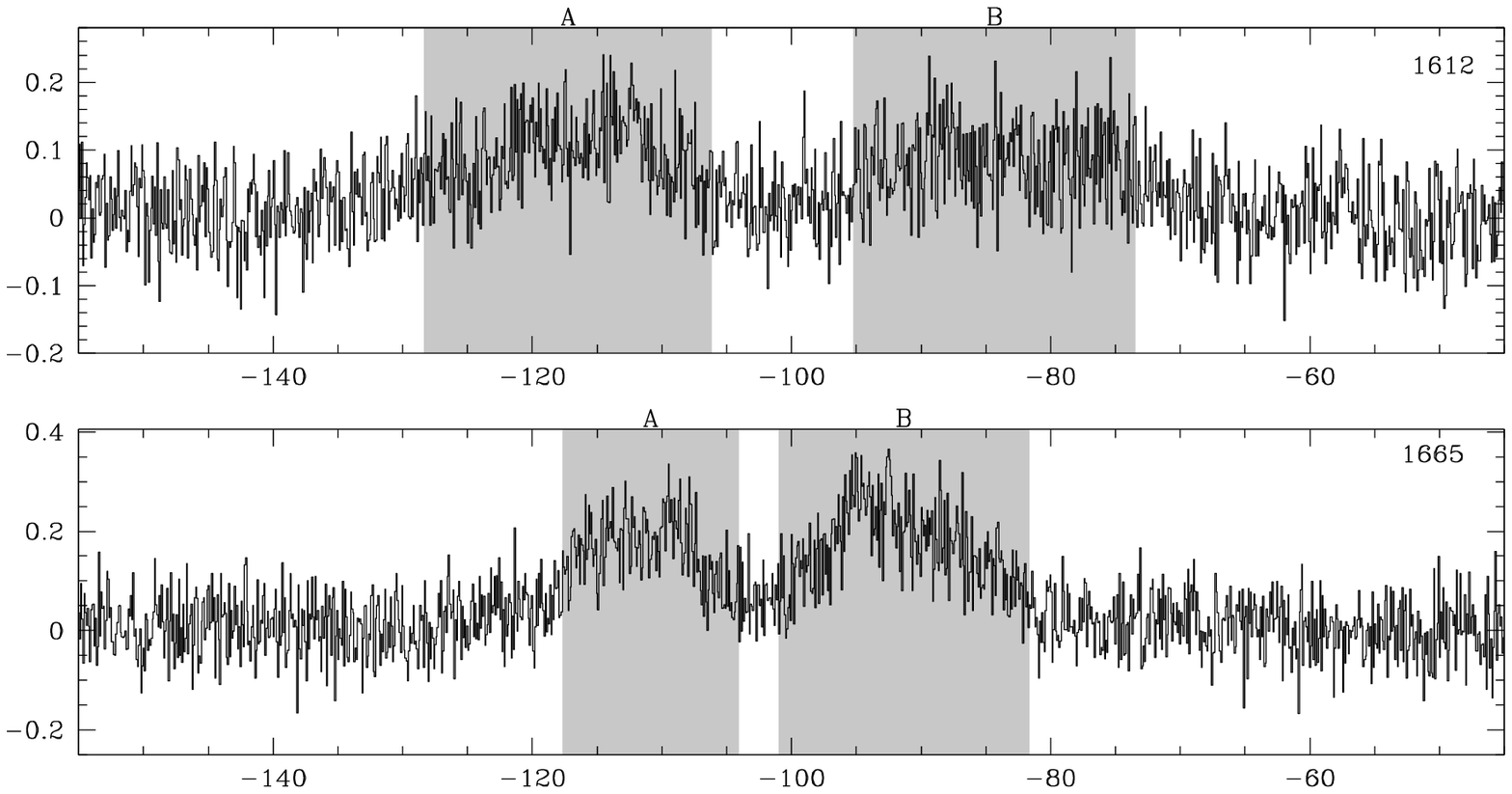}
\includegraphics[width=0.9\textwidth]{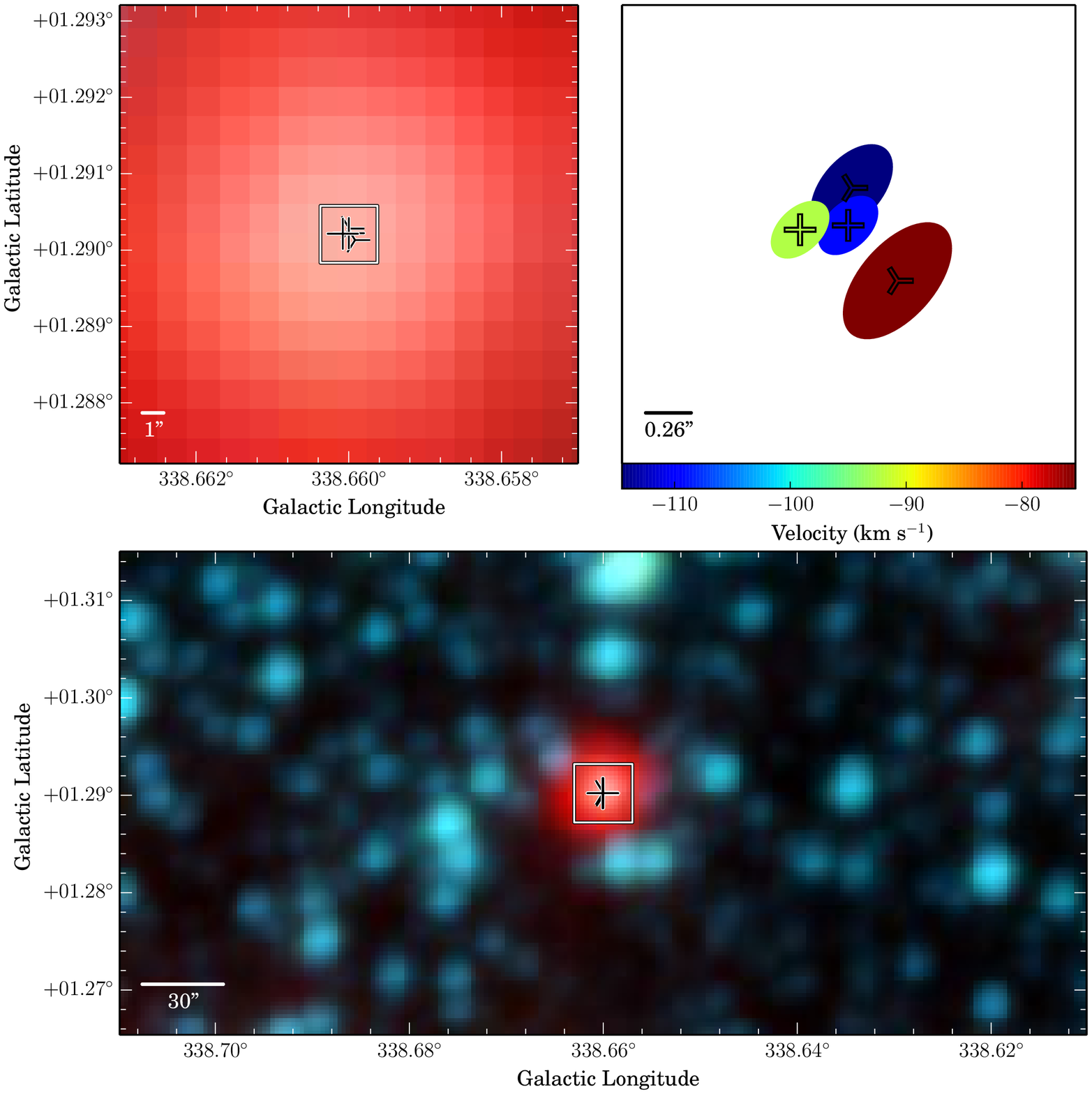}
\caption{G338.660$+$1.290 -- ES}
\label{G338.660}
\end{figure*}

\noindent{\bf G338.660+1.290.} This maser site is identified as an evolved star site. It has 1612 and 1665\,MHz OH maser spots, as shown in Figure \ref{G338.660}. The line-width of these maser spots is quite broad with $\sim$20\kms for the 1612\,MHz transition and $\sim$16\kms for the 1665\,MHz transition. These two masers are detected even on the longest baselines and are not detected in the other lines, thus they are unlikely to be diffuse OH emission. In the WISE three-color image, the star is very red (bright at 12 \um).

\noindent{\bf G338.874$-$0.821.} The unassociated emission in the 1612\,MHz spectrum is from a nearby source, G338.972$-$0.734.% in Figure \ref{figG338.972}

\noindent{\bf G339.282+0.136.} This maser site only exhibits one maser spot at the 1665\,MHz transition. It is also associated with 6.7\,GHz methanol masers and is identified as a star formation site. The line-width of this maser spot is $\sim$4\kms. In the GLIMPSE three-color image, it is associated with a very faint EGO. \citet{Cae2014} reported the variability of this 1665\,MHz maser, with the highest peak decreasing from 1.5 to 0.7 Jy between 2004 and 2005. The 1667\,MHz OH maser they detected was listed in our maser site G339.294+0.139.

\noindent{\bf G339.294+0.139.} This maser site only has one maser spot at 1667\,MHz transition. It is also associated with 6.7\,GHz methanol masers and is thus identified as a star formation site. The line-width of this maser spot is quite narrow, $\sim$0.5\kms. In the GLIMPSE three-color image, it is associated with an EGO. \citet{Cae2014} detected high linear polarization of the 1667\,MHz maser at $-$73.5\kms ($>50$ per\,cent).

\noindent{\bf G339.986$-$0.425.} The unassociated emission in the 1612\,MHz spectrum is from a nearby source, G340.144$-$0.430. %in Figure \ref{figG340.144}

\noindent{\bf G340.043-0.092.} This source is an evolved star. From \citet{Seb1997} observations, it has the typical double-horned profile at 1612\,MHz, which spreads from $-$46.8 to $-$11.8\kms. In our observation, we did not detect the 1612\,MHz OH maser and only detected one red-shifted maser spot at 1667\,MHz, which is $\sim$1.5\kms width.

\noindent{\bf G340.246$-$0.048.} This maser site is identified as an unknown site. It only contains one 1612\,MHz OH maser with line-width of $\sim$1\kms. In the GLIMPSE three-color image, it is located in the 12 \um extended emission (red background) and is not associated with any stellar-like objects. It might be associated with star formation.

\noindent{\bf G341.083$-$1.084.} This maser site is associated with a variable star of Mira Cet type, which is well studied (\citealt{Mce2012}). It only exhibits 1665 and 1667\,MHz OH masers. In the GLIMPSE three-color image, the pixels are saturated.

\begin{figure*}
\includegraphics[width=0.9\textwidth]{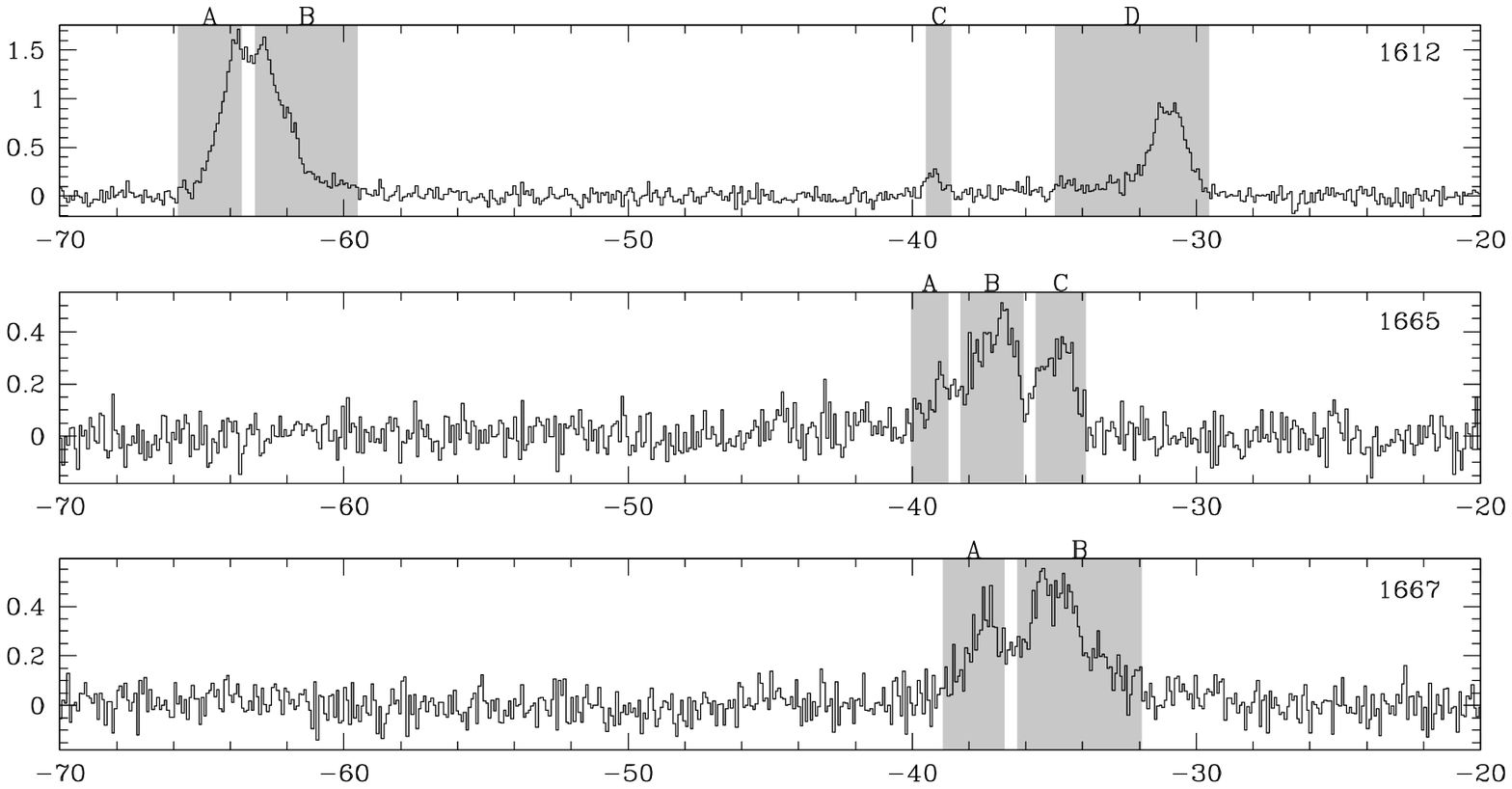}
\includegraphics[width=0.9\textwidth]{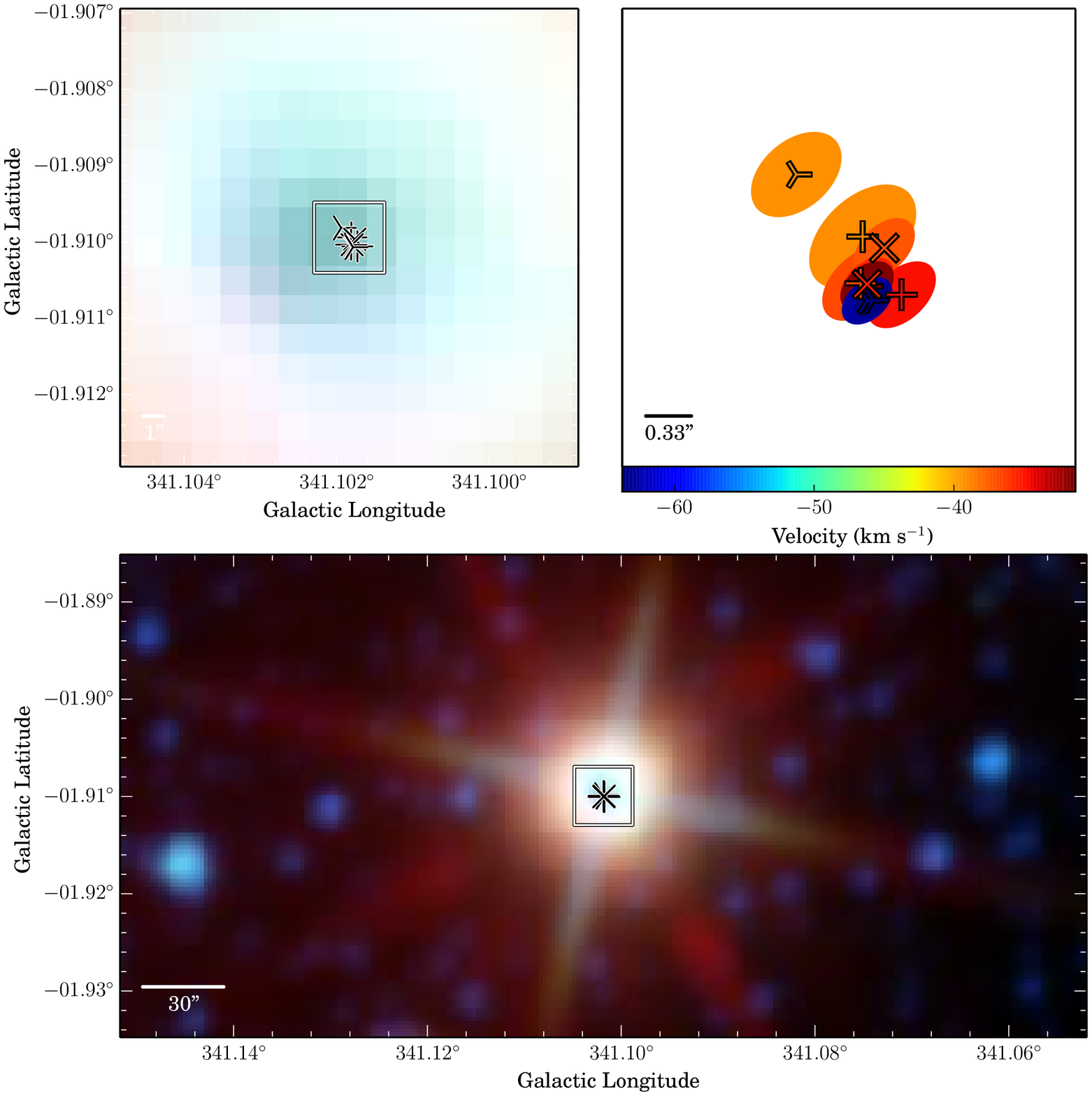}
\caption{G341.102$-$1.910 -- ES}
\label{G341.102}
\end{figure*}

\noindent{\bf G341.102$-$1.910.} This maser site is associated with an evolved star. It not only shows the typical double-horned profile at 1612\,MHz, but also a maser spot between the velocity of the center star and the red-shifted maser spot (named as the middle maser spot), shown in Figure \ref{G341.102}. The five maser spots in 1665 and 1667\,MHz transitions locate between the middle and the red-shifted 1612\,MHz maser spots.

%\noindent{\bf G341.276+0.062.} This maser site has two maser spots at 1665\,MHz transition. It is also associated with 6.7\,GHz methanol masers and is identified as a star formation site. The line-width of this maser spot is $\sim$0.8\kms. In the GLIMPSE three-color image, it is associated with a bright EGO.

\begin{figure*}
\includegraphics[width=0.9\textwidth]{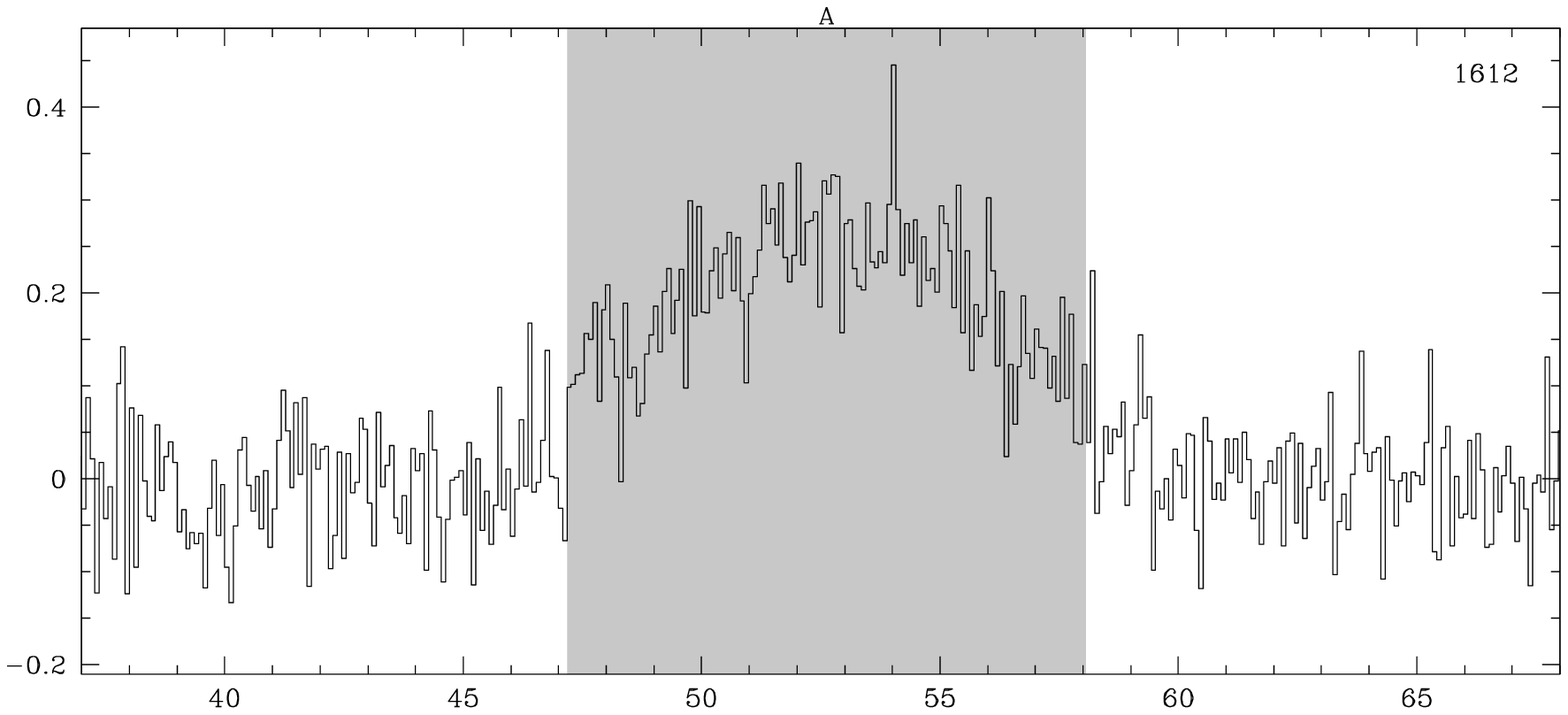}
\includegraphics[width=0.9\textwidth]{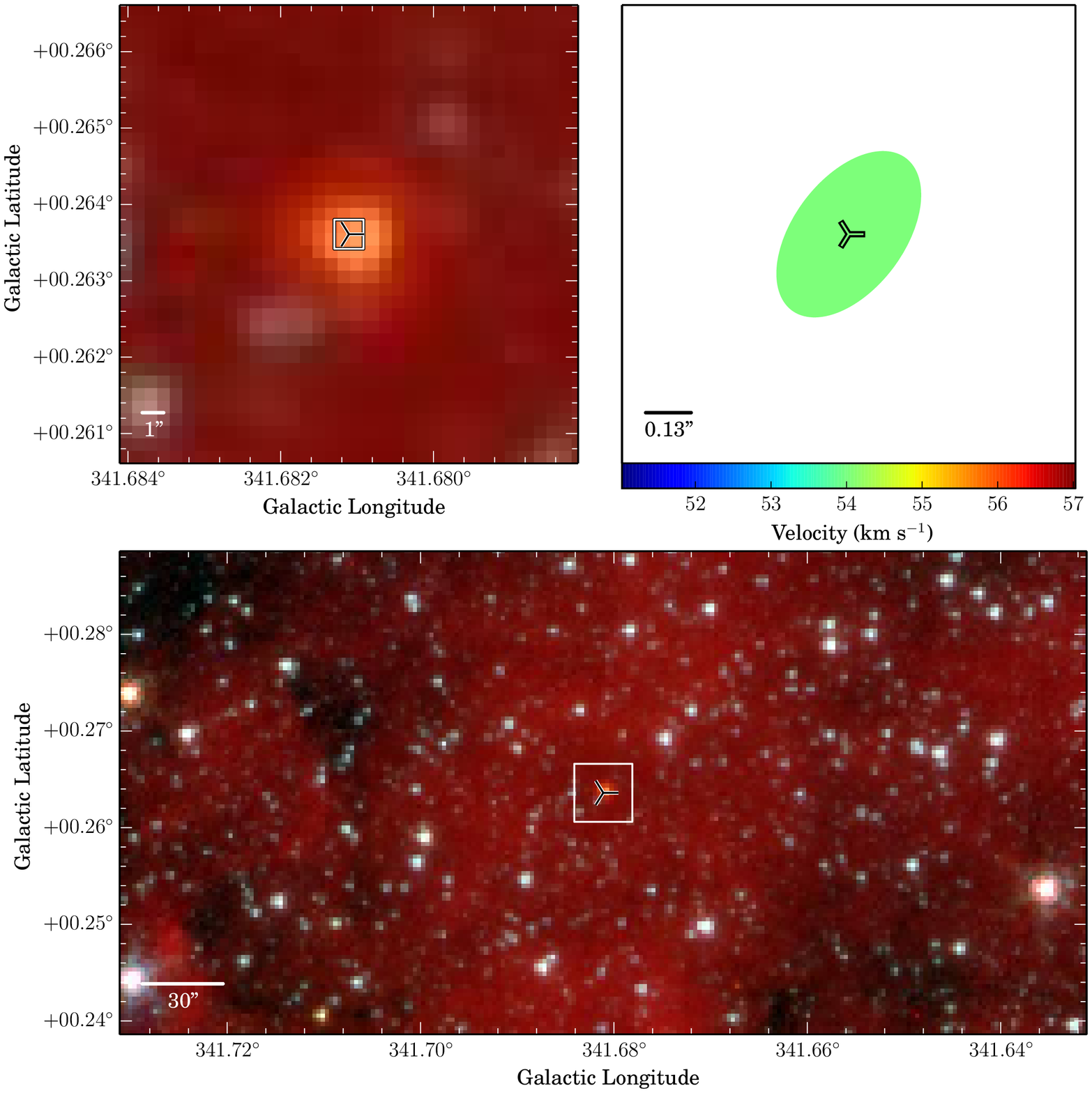}
\caption{G341.681$+$0.264 -- U}
\label{G341.681}
\end{figure*}

\noindent{\bf G341.681+0.264.} This maser site is identified as an unknown site. It only exhibits a very broad 1612\,MHz OH maser spot with line-width of $\sim$12\kms, shown in Figure \ref{G341.681}. This emission is detected on even the longest baselines and is not detected in the other lines, thus it is unlikely to be diffuse OH emission. In the GLIMPSE three-color image, it is associated with a faint stellar-like object, which may be an EGO or an evolved star.

\noindent{\bf G342.004+0.251.} This maser site is identified as an evolved star site. In HOPS, water masers were also observed towards this site with the velocity range from $-$45\kms to $-$20\kms. In our observations, there are three peaks in the 1612\,MHz spectrum and the velocity range is from $-$53\kms to $-$32\kms. The blue-shifted component of the 1612\,MHz OH maser is more blue-shifted than the water masers and its red-shifted component is absent.

\noindent{\bf G342.369+0.140.} This maser site is a star formation site, which also exhibits 6.7\,GHz methanol masers and 22\,GHz water masers. It only contains two 1612\,MHz OH maser spots and is associated with very faint EGOs in the GLIMPSE three-color image.

%\noindent{\bf G342.644+0.114.} This maser site is classified as an unknown site. It only has one 1665\,MHz OH maser, which is about 2\kms broad. In the GLIMPSE three-color image, it is associated with a bright infrared star and may originate from this star. 
\noindent{\bf G342.902$-$0.144.} The unassociated emission in the 1612\,MHz spectrum is from a nearby source, G343.119$-$0.067. % in Figure \ref{figG343.119}

\noindent{\bf G343.119$-$0.067.} This maser site exhibits one strong maser spot at 1612\,MHz. In the GLIMPSE three-color image, this source is associated with extended dark red emission (12 \um) and we classify it as star formation OH maser site. The GLIMPSE image also shows the nearby star formation OH maser site associated with EGOs.

\noindent{\bf G343.721+1.003.} This sources has similar properties as G340.246$-$0.048 in the 1612\,MHz spectrum and the GLIMPSE three-color map. It might be associated with star formation.

\noindent{\bf G343.929+0.125.} This maser site is a star formation site, which also has 6.7\,GHz methanol masers. It has three weak maser spots in the 1665\,MHz transition and is associated with EGOs in the GLIMPSE three-color map. \citet{Cae2014} reported the variability of the strong 1665\,MHz left-hand circular polarization component at 12\kms and the absence of the 1667\,MHz maser.

\noindent{\bf G343.966+0.815.} This maser site is identified as an unknown maser site. It shows a double-horned profile in the 1667\,MHz transition and is associated with an IR stellar-like source in the GLIMPSE three-color map. It may be associated with an evolved star.

\section{Conclusions}
\label{conclusion}

In this paper, we present accurate positions for ground-state OH masers in the SPLASH pilot region, between Galactic longitudes of $334^{\circ}$ and $344^{\circ}$ and Galactic latitudes of $-2^{\circ}$ and $+2^{\circ}$. We also identify the associated astrophysical objects for these maser sites. 

We obtain a total of 215 maser sites, which exhibit maser emission at one, two or three transitions. More than half of these 215 maser sites (111/215) are new detections. Of these 215 maser sites, 154 sites show maser emission at 1612\,MHz, 61 and 57 sites emit maser emission at 1665\,MHz and 1667\,MHz, and 9 sites exhibit the rare 1720\,MHz OH masers.

We identify 122 maser sites associated with evolved stars. These maser sites usually have the characteristic double-horned profile at 1612\,MHz, sometimes accompanied by 1665 and/or 1667\,MHz OH masers. In the GLIMPSE or WISE three-color images, these maser sites are associated with bright IR stellar-like sources. One PN site G336.644$-$0.695 from the evolved star category is detected with OH masers at 1612, 1667 and 1720\,MHz and may represent a very peculiar phase of the PN evolution. 64 maser sites are classified as being associated with star formation. These sites commonly have several strong maser spots at main-line transitions and occasionally also exhibit 1612 or 1720\,MHz OH masers. In the IR images, they tend to be associated with EGOs, which are signatures of star formation. Two 1720\,MHz maser sites are associated with SNRs and trace the interaction of the SNR with the surrounding giant molecular cloud. 27 maser sites are categorised as unknown because of the lack of information from literature and IR images. We find the size of most OH maser sites (98 per\,cent) are smaller than 2.5$\arcsec$ according to their accurate positions. We also find the angular size of OH maser sites associated with evolved stars are smaller than that of star formation sites, which is consistent with 22\,GHz water masers in HOPS. For evolved star sites, OH masers in the 1612\,MHz transition are dominant (98 per\,cent). For star formation sites, main-line transitions have the largest overlap, i.e. 62 per\,cent of 1665\,MHz OH masers are associated with a 1667\,MHz OH maser.

We classify evolved star sites showing the 1612\,MHz transition according to integrated flux densities of blue- and red-shifted components. We find that symmetric sources are redder than asymmetric sources in the WISE [12]-[22] colors. This may be mainly due to a younger age of symmetric sources, compared to asymmetric sources.

We study associations of star formation OH maser sites with 6.7\,GHz methanol masers from the MMB survey and 22\,GHz water masers from HOPS. We find that OH masers have the largest overlap with methanol masers (73 per\,cent), which may be due to the high sensitivity of the methanol maser survey. We also find six of 64 star formation OH maser sites in the SPLASH pilot region are associated with continuum sources at 1.7\,GHz, which is lower than the fraction at 9\,GHz. This lower value is likely due to the frequency we observe, at which \hii regions may be optically thick, or may be caused by the low sensitivity of our observations.

We did not detect any maser emission in 21 fields. From their Parkes spectra, we find they tend to have simpler and weaker profiles. We conclude that about half of non-detections are the result of intrinsic variability.

\acknowledgments The Australia Telescope Compact Array is part of the Australia Telescope which is funded by the Commonwealth of Australia for operation as a National Facility managed by CSIRO. This research has made use of: NASA's Astrophysics Data System Abstract Service; and the SIMBAD data base, operated at CDS, Strasbourg, France. This work is based in part on observations made with the Spitzer Space Telescope, which is operated by the Jet Propulsion Laboratory, California Institute of Technology under a contract with National Aeronautics and Space Administration (NASA). This publication also makes use of data products from the Wide-field Infrared Survey Explorer, which is a joint project of the University of California Institute of Technology, funded by NASA. H.-H.Q. would thank the China Scholarship Council (CSC) support. SLB is a recipient of a L'Or\'eal-UNESCO for Women in Science Fellowship. JFG is partially supported by MINECO (Spain) grant  AYA2014-57369-C3-3 (co-funded by FEDER).
%JFG is partially supported by MICINN (Spain) grants AYA2011-30228-C03-01 and AYA2014-57369-C3-3 (co-funded by FEDER). 
%This work was supported in part by the Major Program of National Natural Science Foundation of China (grants 11590780 and 11590784), National Basic Research Program of China (973 program) No. 2012CB821806, the Strategic Priority Research Program ``The Emergence of Cosmological Structures'' of the Chinese Academy of Sciences, Grant No. XDB09000000.

\label{lastpage}


\begin{thebibliography}{99}

\bibitem[Argon et al.(2000)]
{Are2000} Argon, A.~L., Reid, M.~J., \& Menten, K.~M.\ 2000, \apjs, 129, 159 

\bibitem[Baan et al.(1982)]
{Bae1982} Baan, W.~A., Wood, P.~A.~D., \& Haschick, A.~D.\ 1982, \apjl, 260, L49 

\bibitem[Bains et al.(2009)]
{Bae2009} Bains, I., Cohen, M., Chapman, J.~M., Deacon, R.~M., \& Redman, M.~P.\ 2009, \mnras, 397, 1386 

\bibitem[Benjamin et al.(2003)]
{Bee2003} Benjamin, R.~A., Churchwell, E., Babler, B.~L., et al.\ 2003, \pasp, 115, 953

\bibitem[Blum et al.(2006)]
{Ble2006} Blum, R.~D., Mould, J.~R., Olsen, K.~A., et al.\ 2006, \aj, 132, 2034 

\bibitem[Breen et al.(2010a)]
{Bre2010a}Breen, S.~L., Ellingsen, S.~P., Caswell, J.~L., \& Lewis, B.~E.\ 2010a, \mnras, 401, 2219 

\bibitem[Breen et al.(2010b)]
{Bre2010b}Breen, S.~L., Caswell, J.~L., Ellingsen, S.~P., \& Phillips, C.~J.\ 2010b, \mnras, 406, 1487 

%\bibitem[Breen et al.(2013)]
%{Bre2013} Breen, S.~L., Ellingsen, S.~P., Contreras, Y., et al.\ 2013, \mnras, 435, 524 

\bibitem[Caswell et al.(1980)]
{CH1980} Caswell, J.~L., Haynes, R.~F., \& Goss, W.~M.\ 1980, AuJPh, 33, 639 

\bibitem[Caswell \& Haynes(1983a)]
{CH1983a} Caswell, J.~L., \& Haynes, R.~F.\ 1983a, AuJPh, 36, 361 

\bibitem[Caswell \& Haynes(1983b)]
{CH1983b} Caswell, J.~L., \& Haynes, R.~F.\ 1983b, AuJPh, 36, 417 

\bibitem[Caswell \& Haynes(1987)]
{CH1987} Caswell, J.~L., \& Haynes, R.~F.\ 1987, AuJPh, 40, 215 

\bibitem[Caswell(1998)]
{Ca1998} Caswell, J.~L.\ 1998, \mnras, 297, 215

\bibitem[Caswell(2004)]
{Ca2004} Caswell, J.~L.\ 2004, \mnras, 352, 101 

%\bibitem[Caswell et al.(2010)]
%{Cae2010} Caswell, J.~L., Fuller, G.~A., Green, J.~A., et al.\ 2010, \mnras, 404, 1029 

\bibitem[Caswell et al.(2011)]
{Cae2011} Caswell, J.~L., Fuller, G.~A., Green, J.~A., et al.\ 2011, \mnras, 417, 1964 

\bibitem[Caswell et al.(2014)]
{Cae2014} Caswell, J.~L., Green, J.~A., \& Phillips, C.~J.\ 2014, \mnras, 439, 1680 

\bibitem[Chen \& Yang(2012)]
{Che2012} Chen, P.~S., \& Yang, X.~H.\ 2012, \aj, 144, 104 

\bibitem[Chen et al.(2009)]
{Che2009} Chen, X., Ellingsen, S.~P., \& Shen, Z.-Q.\ 2009, \mnras, 396, 1603 

\bibitem[Churchwell et al.(2009)]
{Chu2009} Churchwell, E., Babler, B.~L., Meade, M.~R., et al.\ 2009, \pasp, 121, 213 

\bibitem[Clegg \& Cordes(1991)]
{CJ1991} Clegg, A.~W., \& Cordes, J.~M.\ 1991, \apj, 374, 150 

\bibitem[Cyganowski et al.(2008)]
{Cye2008} Cyganowski, C.~J., Whitney, B.~A., Holden, E., et al.\ 2008, \aj, 136, 2391-2412

\bibitem[Dawson et al.(2014)]
{Dae2014} Dawson, J.~R., Walsh, A.~J., Jones, P.~A., et al.\ 2014, \mnras, 439, 1596 

\bibitem[Edris et al.(2007)]
{Ede2007} Edris, K.~A., Fuller, G.~A., \& Cohen, R.~J.\ 2007, \aap, 465, 865

\bibitem[Forster \& Caswell(1989)]
{FC1989} Forster, J.~R., \& Caswell, J.~L.\ 1989, \aap, 213, 339 

\bibitem[Forster \& Caswell(2000)]
{FC2000} Forster, J.~R., \& Caswell, J.~L.\ 2000, \apj, 530, 371 

\bibitem[Frail et al.(1996)]
{Fre1996} Frail, D.~A., Goss, W.~M., Reynoso, E.~M., et al.\ 1996, \aj, 111, 1651 

\bibitem[G{\'e}rard et al.(1998)]
{Gee1998} G{\'e}rard, E., Crovisier, J., Colom, P., et al.\ 1998, \planss, 46, 569 

\bibitem[Goss \& Robinson(1968)]
{Goe1968} Goss, W.~M., \& Robinson, B.~J.\ 1968, \aplett, 2, 81

\bibitem[Greaves(2008)]
{Gre2008}Greaves, J.\ 2008, PZP, 8, 

%\bibitem[Green et al.(2010)]
%{Gre2010} Green, J.~A., Caswell, J.~L., Fuller, G.~A., et al.\ 2010, \mnras, 409, 913 

\bibitem[Green et al.(2012)]
{Gre2012a} Green, J.~A., McClure-Griffiths, N.~M., Caswell, J.~L., Robishaw, T., \& Harvey-Smith, L.\ 2012, \mnras, 425, 2530 

%\bibitem[Green et al.(2012b)]
%{Gre2012b} Green, J.~A., Caswell, J.~L., Fuller, G.~A., et al.\ 2012, \mnras, 420, 3108


\bibitem[Gundermann(1965)]
{Gu1965} Gundermann, E.~J.\ 1965, Ph.D.~Thesis,

%\bibitem[Jiang et al.(2010)]
%{Jie2010} Jiang, B., Chen, Y., Wang, J., et al.\ 2010, \apj, 712, 1147

\bibitem[Jones et al.(1982)]
{Joe1982} Jones, T.~J., Hyland, A.~R., Gatley, I., \& Caswell, J.~L.\ 1982, \apj, 253, 208

\bibitem[Kurtz et al.(1994)]
{Kue1994} Kurtz, S., Churchwell, E., \& Wood, D.~O.~S.\ 1994, \apjs, 91, 659

\bibitem[Lewis et al.(1995)]
{Lee1995} Lewis, B.~M., David, P., \& Le Squeren, A.~M.\ 1995, \aaps, 111, 237 

%\bibitem[Likkel(1989)]
%{Lik1989} Likkel, L.\ 1989, \apj, 344, 350

%\bibitem[Lo \& Bechis(1973)]
%{Loe1973} Lo, K.~Y., \& Bechis, K.~P.\ 1973, \apjl, 185, L71

\bibitem[Lumsden et al.(2013)]
{Lue2013} Lumsden, S.~L., Hoare, M.~G., Urquhart, J.~S., et al.\ 2013, \apjs, 208, 11 

\bibitem[McDonald et al.(2012)]
{Mce2012} McDonald, I., Zijlstra, A.~A., \& Boyer, M.~L.\ 2012, \mnras, 427, 343 

%\bibitem[Nakashima \& Deguchi(2003)]
%{Nae2003} Nakashima, J.-I., \& Deguchi, S.\ 2003, \pasj, 55, 229

\bibitem[Nguyen-Q-Rieu et al.(1979)]
{Nge1979} Nguyen-Q-Rieu, Laury-Micoulaut, C., Winnberg, A., \& Schultz, G.~V.\ 1979, \aap, 75, 351 

\bibitem[Qiao et al.(2014)]
{Qie2014} Qiao, H., Li, J., Shen, Z., Chen, X., \& Zheng, X.\ 2014, \mnras, 441, 3137 

\bibitem[Qiao et al.(2016)]
{Qie2016} Qiao, H.-H., Walsh, A.~J., Gomez, J.~F., et al. \ 2016, \apj, 817, 37

\bibitem[Reid \& Moran(1981)]
{Ree1981} Reid, M.~J., \& Moran, J.~M.\ 1981, \araa, 19, 231 

\bibitem[Reid et al.(1988)]
{Ree1988} Reid, M.~J., Schneps, M.~H., Moran, J.~M., et al.\ 1988, \apj, 330, 809 

\bibitem[Reid(2002)]
{Rei2002} Reid, M.~J.\ 2002, IAUS, 206, 506 

\bibitem[Robitaille et al.(2008)]
{Roe2008} Robitaille, T.~P., Meade, M.~R., Babler, B.~L., et al.\ 2008, \aj, 136, 2413 

\bibitem[Sault et al.(1995)]
{Sae1995} Sault, R.~J., Teuben, P.~J., \& Wright, M.~C.~H.\ 1995, ASPC, 77, 433 

\bibitem[Sevenster et al.(1997a)]
{Sea1997} Sevenster, M.~N., Chapman, J.~M., Habing, H.~J., Killeen, N.~E.~B., \& Lindqvist, M.\ 1997, \aaps, 122, 79 

\bibitem[Sevenster et al.(1997b)]
{Seb1997} Sevenster, M.~N., Chapman, J.~M., Habing, H.~J., Killeen, N.~E.~B., \& Lindqvist, M.\ 1997, \aaps, 124, 509

\bibitem[Sevenster et al.(2001a)]
{Sea2001} Sevenster, M.~N., van Langevelde, H.~J., Moody, R.~A., et al.\ 2001, \aap, 366, 481 

\bibitem[Sevenster \& Chapman(2001b)]
{Seb2001} Sevenster, M.~N., \& Chapman, J.~M.\ 2001, \apjl, 546, L119 

\bibitem[Smith(2003)]
{Smi2003} Smith, B.~J.\ 2003, \aj, 126, 935 

%\bibitem[Stupar \& Parker(2011)]
%{Ste2011} Stupar, M., \& Parker, Q.~A.\ 2011, \mnras, 414, 2282

\bibitem[te Lintel Hekkert et al.(1989)]
{Tee1989} te Lintel Hekkert, P., Versteege-Hensel, H.~A., Habing, H.~J., \& Wiertz, M.\ 1989, \aaps, 78, 399

\bibitem[te Lintel Hekkert et al.(1991)]
{Tee1991} te Lintel Hekkert, P., Caswell, J.~L., Habing, H.~J., et al.\ 1991, \aaps, 90, 327

\bibitem[Titmarsh et al.(2014)]
{Tie2014} Titmarsh, A.~M., Ellingsen, S.~P., Breen, S.~L., Caswell, J.~L., \& Voronkov, M.~A.\ 2014, \mnras, 443, 2923 

\bibitem[Titmarsh et al.(2016)]
{Tie2016} Titmarsh, A.~M., Ellingsen, S.~P., Breen, S.~L., Caswell, J.~L., \& Voronkov, M.~A.\ 2016, \mnras, 459, 157 

\bibitem[Uscanga et al.(2012)]
{Use2012} Uscanga, L., G{\'o}mez, J.~F., Su{\'a}rez, O., \& Miranda, L.~F.\ 2012, \aap, 547, A40 

\bibitem[Uscanga et al.(2014)]
{Use2014} Uscanga, L., G{\'o}mez, J.~F., Miranda, L.~F., et al.\ 2014, \mnras, 444, 217 

\bibitem[van de Steene \& Pottasch(1993)]
{Vae1993} van de Steene, G.~C.~M., \& Pottasch, S.~R.\ 1993, \aap, 274, 895

\bibitem[Walsh et al.(1998)]
{Wae1998} Walsh, A.~J., Burton, M.~G., Hyland, A.~R., \& Robinson, G.\ 1998, \mnras, 301, 640 

\bibitem[Walsh et al.(2012)]
{Wae2012} Walsh, A.~J., Purcell, C., Longmore, S., Jordan, C.~H., \& Lowe, V.\ 2012, \pasa, 29, 262

\bibitem[Walsh et al.(2014)]
{Wae2014} Walsh, A.~J., Purcell, C.~R., Longmore, S.~N., et al.\ 2014, \mnras, 442, 2240 

\bibitem[Weaver et al.(1965)]
{Wee1965} Weaver, H., Williams, D.~R.~W., Dieter, N.~H., \& Lum, W.~T.\ 1965, \nat, 208, 29 

\bibitem[Wright et al.(2010)]
{Wre2010} Wright, E.~L., Eisenhardt, P.~R.~M., Mainzer, A.~K., et al.\ 2010, \aj, 140, 1868-1881

\end{thebibliography}
\end{document}